\documentclass[11pt,a4paper]{article}
\usepackage{jheppub}

\preprint{LU TP 15-07, MCnet-15-02}

\usepackage{epstopdf} 
\usepackage{inputenc}
\usepackage{xspace}
\usepackage{amsfonts}
\usepackage{url}
\inputencoding{latin1}
\usepackage{def}

\allowdisplaybreaks

\title{Recursion in multiplet bases for tree-level MHV gluon amplitudes}

\author[a,1]{Yi-Jian Du
\note{On leave from Center for Field Theory and Particle Physics, Department of Physics, Fudan University.}}
\author[a]{Malin Sj\"odahl}
\author[a]{and Johan Thorén}

\affiliation[a]{Department of Astronomy and Theoretical Physics, Lund University,\\
  S\"olvegatan 14A, 223\,62~Lund, Sweden}

\emailAdd{yijian.du@thep.lu.se}
\emailAdd{malin.sjodahl@thep.lu.se}
\emailAdd{johan.thoren@thep.lu.se}

\date{\today}
\abstract{
We investigate the construction of tree-level MHV gluon
amplitudes in multiplet bases using BCFW recursion.
The multiplet basis decomposition can either be obtained by
decomposing results derived in (for example) the DDM basis
or by formulating the recursion directly in the multiplet basis.
We focus on the latter approach and show how to efficiently deal
with the color structure appearing in the recursion.
For illustration, we also explicitly calculate the four-, five- and six-gluon amplitudes.
}


\begin{document}
\maketitle

\section{Introduction}

Scattering amplitudes are essential tools for understanding
multi-parton processes in high energy physics.
Traditional calculations based on Feynman diagrams have proven
extremely cumbersome due to the large number of diagrams and
the correlation of the color and kinematic factors. To simplify the
calculations of amplitudes, large efforts have been made,
including the development of the spinor helicity formalism
\cite{Berends:1981rb, Causmaecker:1981bg, Kleiss:1985yh, Xu:1986xb, Gunion:1985vca},
the Berends-Giele recursion for off-shell currents
\cite{Berends:1987me},
various forms of color decompositions
\cite{Cvi76,Cvitanovic:1980bu,Dittner:1972hm, Zeppenfeld:1988bz, Paton:1969je, Berends:1987cv, Mangano:1987xk, Mangano:1988kk, Kosower:1988kh, Nagy:2007ty, Sjodahl:2009wx, Alwall:2011uj, Sjodahl:2014opa, Platzer:2012np, DelDuca:1999ha,DelDuca:1999rs,Kanaki:2000ms,Maltoni:2002mq,Keppeler:2012ih, Sjodahl:2015qoa,Kol:2014yua},
the Parke-Taylor formula for color-ordered maximally helicity violating
(MHV) amplitudes \cite{Parke:1986gb,Berends:1987me},
the Kleiss-Kuijf (KK) relation \cite{Kleiss:1988ne},
the twistor string method \cite{Witten:2003nn}, Cachazo-Svrcek-Witten (CSW) rules
\cite{Cachazo:2004kj},
Britto-Cachazo-Feng-Witten (BCFW) recursion
\cite{Britto:2004ap,Britto:2005fq}
and the Bern-Carrasco-Johansson (BCJ) relation \cite{Bern:2008qj} between color-ordered amplitudes.
Reviews of related topics can be found in \cite{Mangano:1990by,Dixon:1996wi,Cachazo:2005ga,Bern:2007dw,Feng:2011np,Peskin:2011in,Benincasa:2013faa,Ellis:2011cr,Elvang:2013cua}.

 In most of this work the color decomposition plays an important
role for understanding and simplifying scattering amplitudes.
Several types of color decompositions for gluon amplitudes
are available, most notably the trace basis decomposition
\cite{Paton:1969je, Berends:1987cv, Mangano:1987xk, Mangano:1988kk, Kosower:1988kh, Nagy:2007ty, Sjodahl:2009wx, Alwall:2011uj, Sjodahl:2014opa, Platzer:2012np},
the Del~Duca-Dixon-Maltoni (DDM) basis decomposition \cite{DelDuca:1999ha, DelDuca:1999rs}
and the color-flow basis decomposition \cite{Maltoni:2002mq,Kanaki:2000ms}.
Since recently it is also known how to construct orthogonal
group theory based multiplet bases for any number of
quarks and gluons \cite{Keppeler:2012ih}.

\begin{figure}
  \centering
\begin{equation*}
\raisebox{-0.315\height}{
	\includegraphics[scale=0.6]{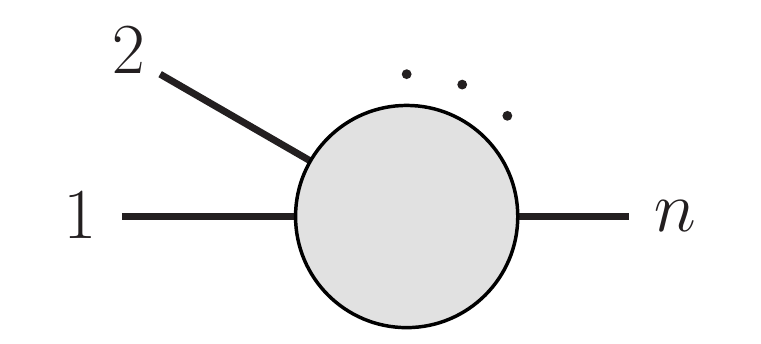}
}
\hspace{-2mm}
=
\sum_{i=2}^{n-2}{
	\sum_{h=\pm}{
	\hspace{-2mm}
		\raisebox{-0.27\height}{
			\includegraphics[scale=0.6]{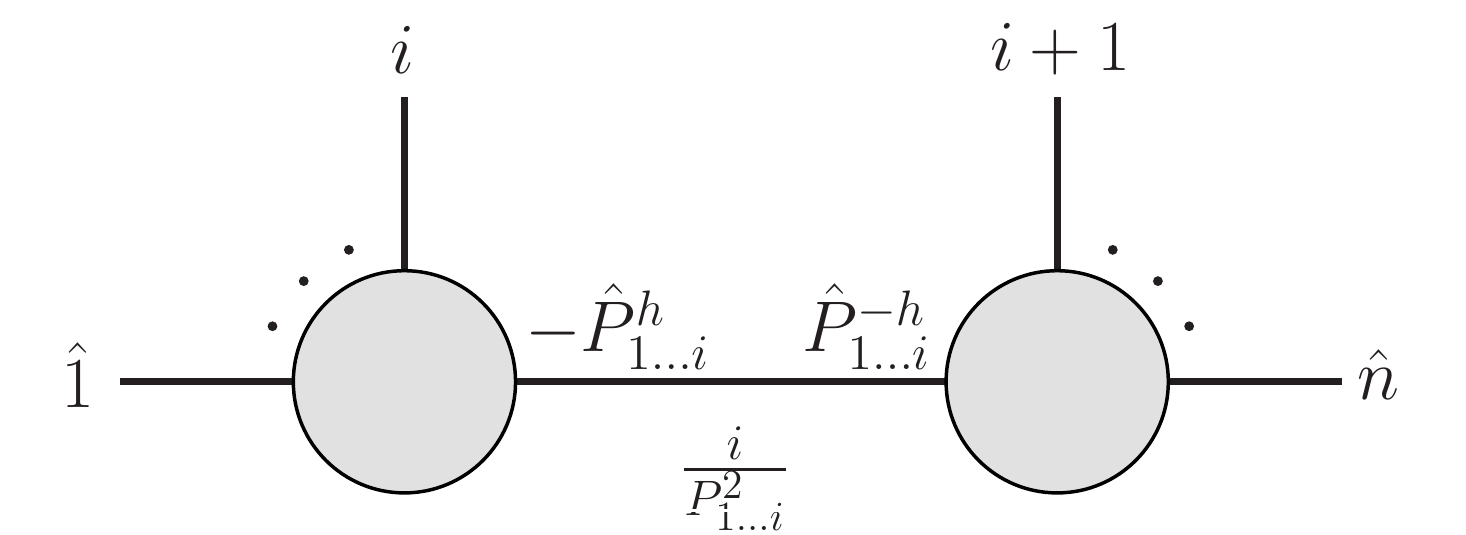}
		}
	}
}
\end{equation*}
 \caption{BCFW recursion for color-ordered tree-level gluon amplitudes.} \label{BCFW}
\end{figure}

While the topic of scattering amplitude recursion relations
has been explored for the kinematic factors (the color-ordered amplitudes) of the first three bases for a while
\cite{Berends:1987me,Britto:2004ap,Britto:2005fq},
this field is unknown territory in the context of multiplet bases.
In the present paper we take the first step to remedy this
by showing how to use BCFW recursion relations for multiplet bases
in the case of MHV amplitudes in pure Yang-Mills theory.
To set the stage for this task, we first give a brief overview of
the standard color decompositions and the present status of
recursion strategies.
\begin{itemize}
\item {\bf Trace bases}\\
 The color decomposition for a tree-level gluon amplitude
with $\Ng$ gluons in trace bases is given
by\footnote{In the color decomposition formulae, we suppress the helicity
of the external legs for convenience.
Only when discussing amplitudes with a particular helicity
configuration, we specify the helicity information of the external legs.}
\cite{Paton:1969je, Berends:1987cv, Mangano:1987xk, Mangano:1988kk, Kosower:1988kh, Nagy:2007ty, Sjodahl:2009wx, Alwall:2011uj, Sjodahl:2014opa},
\bea
\mathcal{M}(g_1,g_2,\dots, g_n)=
g^{n-2}\left(1\over\sqrt{\TR}\right)^n\Sl_{\sigma,\mbox{ \scriptsize s.t. } \sigma_1=1}
\Tr(t^{g_1}t^{g_{\sigma_2}}\dots t^{g_{\sigma_n}})A(\sigma),~~\Label{Trace-decomposition}
\eea
where we have used  a general generator normalization $\tr(t^a t^b)=\TR \delta^{a\,b}$.
The cyclicity of the trace allows for fixing $\sigma_1=1$, thus
leaving $(\Ng-1)!$ color structures in the sum. The kinematic
factors, $A(\sigma)$ are called color-ordered amplitudes and
can be calculated from the color-ordered
Feynman rules, also known as the color-stripped Feynman rules.
These bases (spanning sets) are easily extended to loop level,
upon which products of traces occur, \cite{Bern:1990ux, Nagy:2007ty, Sjodahl:2009wx, Sjodahl:2014opa},
as well as to processes with quarks,
requiring open quark-lines in addition to the traces \cite{Kosower:1988kh,Mangano:1988kk,Mangano:1990by,Nagy:2007ty,Sjodahl:2009wx,Sjodahl:2014opa}.

\item {\bf Color-flow bases}\\
An approach similar to the trace basis approach is given by the color-flow
bases \cite{Maltoni:2002mq,Kanaki:2000ms}. Here the gluon field is
rewritten in terms of the fundamental representation $(A_\mu)^i_j$,
$i,j=1,..,\Nc$, for $\Nc$ colors,
and the color structure is described in terms of flow of color.
For tree-level gluon amplitudes the color decomposition is given by
\bea
\mathcal{M}(g_1,g_2,\dots, g_n)={g^{n-2}\left(1\over\sqrt{\TR}\right)^n}
\Sl_{\sigma,\mbox{ \scriptsize s.t. } \sigma_1=1}
\delta^{i_{\sigma_1}}_{j_{\sigma_2}}\delta^{i_{\sigma_2}}_{j_{\sigma_3}}\dots \delta^{i_{\sigma_n}}_{j_{\sigma_1}}A(\sigma),
~~\Label{Color-flow-decomposition}
\eea
where the sum runs over the $(\Ng-1)!$ permutations from
connecting color lines. It is not hard to argue that
the amplitudes  $A(\sigma)$ are the same as in the trace bases.
Similar to the trace bases, these bases are extendable to processes
at higher order. 
Their advantage lies in better scaling properties for Monte Carlo
treatment of the color structure \cite{Maltoni:2002mq}.

\item {\bf Del Duca-Dixon-Maltoni bases}\\
Tree-level gluon amplitudes may alternatively be decomposed using the
Del Duca-Dixon-Maltoni (DDM) bases \cite{DelDuca:1999ha,DelDuca:1999rs}
\bea
\mathcal{M}({g_1},{g_2},\dots, {g_n})=g^{n-2}\left(1\over\sqrt{\TR}\right)^{n-2}\Sl_{\sigma,\mbox{ \scriptsize s.t. }\sigma_1=1,\,\sigma_n=n}if^{g_{\sigma_1}g_{\sigma_2}i_1} if^{i_1g_{\sigma_3}i_2}\dots i f^{i_{n-3}g_{\sigma_{n-1}}g_{\sigma_n}}A(\sigma),~~\Label{DDM-decomposition}
\eea
where $\sigma_1=1$ and $\sigma_n=n$ are fixed. The color-ordered amplitudes $A(\sigma_1=1,\sigma_2,\dots,\sigma_{n-1},\sigma_n=n)$ form the so called
Kleiss-Kuijf (KK) basis \cite{Kleiss:1988ne}.
All other color-ordered amplitudes, i.e., all the amplitudes in
\eqref{Trace-decomposition} with $\sigma(n)\ne n$,
can be expressed  in the KK basis using the Kleiss-Kuijf (KK) relation
\bea
A(1,\{\alpha\},n,\{\beta\})=(-1)^{n_{\beta}}\Sl_{\sigma\in OP(\alpha\bigcup\beta^T)}A(1,\sigma,n),\Label{KK-relation}
\eea
where
$n_{\beta}$ denotes the number of indices in the set $\beta$, and
the sum runs over all possible permutations which keep
the relative order of indices in the index set $\alpha$
and reverse the relative index order in the index set $\beta$,
while allowing for all possible relative orderings between each
$\alpha_i$ with respect to each $\beta_j$.

Compared to the trace bases and color-flow bases there is clearly
an advantage in needing only $(\Ng-2)!$ rather than $(\Ng-1)!$
spanning vectors. The color decomposition in DDM bases can also be
extended to one-loop level and to amplitudes
containing a quark-antiquark pair \cite{DelDuca:1999rs}.
A proof of the KK relation can be found in \cite{DelDuca:1999rs},
the BCFW approach to the KK relation is presented in \cite{Feng:2010my},
and the Berends-Giele recursion approach for the off-shell KK relation was given in \cite{Fu:2012uy}.
\end{itemize}
In the three color decompositions above, the kinematic factors
can be expressed using the color-ordered
amplitudes that were defined in the trace bases decomposition,
\eqref{Trace-decomposition}.
Employing BCFW recursion, one can express
these amplitudes in terms of products of lower point on-shell
color-ordered amplitudes. Specifically, for the $n$-gluon
color-ordered tree amplitude $A(1,2,3,\dots,n)$, if we shift the
momenta of gluon $1$ and $n$ using some  complex four-vector $q$ and
some complex variable $z$,
\bea
\WH p_1(z)=p_1-zq,~~\WH p_n(z)=p_n+zq,~~\Label{(1,n)-shift}
\eea
such that the gluons $\hat 1$ and $\hat n$ remain on-shell, the color-ordered amplitude $A(1,2,3,\dots,n)$  is given by the BCFW recursion (see \figref{BCFW})
\bea
&&A\left(1,2,\dots,n\right)\nn
&=&\Sl_{i=2}^{n-2}\Sl_{h=\pm}A\left(\WH 1(z_i),2,\dots,i,-\WH P^h_{\WH 1(z_i),2,\dots,i}(z_i)\right){i\over P^2_{ 1,2,\dots,i}}A\left(\WH P^{-h}_{\WH 1(z_i),2,\dots,i},i+1,\dots,n-1,\WH n(z_i)\right),
\eea
where, $P_{ 1,2,\dots,i}=\Sl_{k=1}^ip_k$. For a given $i$, $z_i$ can be solved from

\bea
\WH P^2_{\WH 1,2,\dots,i}(z_i)=\left(p_1+p_2+\dots+p_i-z_i q\right)^2=0.
\eea

While the trace bases, the color-flow bases and the DDM bases
are conceptually simple and well-established in the
literature, they all suffer from being non-orthogonal (unless $\Nc\to \infty$), and
-- in an $\Nc=3$ sense -- non-minimal.
The bases being non-orthogonal implies that the squaring of tree-level
color structures involves $(\Ng-1)!^2$ terms in the case of the trace
and color-flow bases, and $(\Ng-2)!^2$ terms for the pure Yang-Mills
specific DDM bases.
Going to arbitrary order in perturbation theory, new color structures
appear and the number of vectors (neglecting charge conjugation invariance)
increases up to $\mbox{subfactorial}(\Ng)\approx(\Ng!/e)$
in the trace bases \cite{Keppeler:2012ih}.
For practical purposes this means that it is hard to treat the
color structure of processes involving more than $\sim 8$  gluons plus
$\qqbar$-pairs exactly, and currently the most efficient technique for
multi-parton calculations is probably to sample color structures in
the color-flow basis by Monte Carlo techniques \cite{Maltoni:2002mq}.

One way to cure the bad scaling involved in the squaring step
is to use orthogonal bases. In the case of few partons,
such bases, based on the decomposition of the color structure into
irreducible representations, have been around for a while
\cite{Sotiropoulos:1993rd,Kidonakis:1998nf,Kyrieleis:2005dt,
  Dokshitzer:2005ig,Sjodahl:2008fz,Beneke:2009rj},
but a general strategy for basis construction based on multiplets
was presented only recently \cite{Keppeler:2012ih}.
Below we give an overview of their key properties.

\begin{itemize}
\item {\bf Multiplet bases}\\
These bases are based on subgrouping sets of partons and forcing
the parton sets to transform under irreducible representations of SU($\Nc$).
By applying the same subgrouping procedure to all basis vectors,
orthogonal  basis vectors are obtained.

The decomposition into these bases (to any order in perturbation theory)
may be written
\bea
\mathcal{M}(g_1,g_2,\dots,g_n)
\sim\Sl_{\alpha}\Vec^{\alpha}
A^\alpha,~~\Label{Multiplet-decomposition1}
\eea
where $\alpha$ is some (collective) index of the basis vectors
describing the involved representations
and $ A^\alpha$ are the kinematic factors, clearly not equal
to the $A(\sigma)$ in eqs.~(\ref{Trace-decomposition}-\ref{DDM-decomposition}).

As the multiplet bases have no direct connection to tree-level
color structure, they do not typically span a minimal set
for tree-level color structure alone. Instead they are applicable
to any order, and are (or can trivially be made) minimal for any
finite $\Nc$, leading to a significant reduction
in dimension for a large number of partons (see \tabref{tab:RadiationMatrixNonZero}).

\end{itemize}

As the multiplet basis vectors $\Vec^{\alpha}$ do not (generally)
have any cyclic symmetry under exchange of gluon indices,
one can not expect recursion relations as simple as for the trace
basis to hold. On the other hand, being orthogonal, multiplet bases
speed up the squaring of amplitudes very significantly.
We thus expect to gain in the squaring step, at the expense of
a more intricate color decomposition.
This decomposition -- in a recursion context -- is the topic of
the present paper, and we will discuss two different ways of achieving
it.

\begin{itemize}
\item
Clearly one strategy is to express the basis vectors in any of the
bases where the recursion is known in terms of the relevant
multiplet basis.
While this strategy -- which in principle is nothing but a change of basis --
is straight forward, the gain in computation time is unclear
as it involves decomposing $(\Ng-1)!$ or $(\Ng-2)!$ color
structures into an exponentially growing number of basis vectors.
In theory, the exponential times the factorial does scale better than
the factorial square involved in squaring amplitudes in the
non-orthogonal standard bases.
In reality, however, this difference
only becomes significant for a relatively large number of gluons,
(cf. \tabref{tab:RadiationMatrixNonZero}).
Directly rewriting recursion results obtained in other bases may,
however, still be beneficial if one color factor (for example
one trace or one DDM color factor) can be rewritten as a linear
combination of a small number of multiplet basis vectors,
and the non-vanishing projections can be identified quickly.
For tree-level gluon amplitudes, the best option is likely
to use the smaller DDM basis \eqref{DDM-decomposition}.

In principle, for tree-level processes with only gluons it is
also possible to use the Bern-Carrasco-Johansson
relation \cite{Bern:2008qj} (the BCJ relation has been proved in both string theory \cite{BjerrumBohr:2009rd,Stieberger:2009hq} and field theory \cite{Feng:2010my,Tye:2010kg,Chen:2011jxa} ) to reduce the color-ordered amplitudes
in a smaller basis of $(\Ng-3)!$ color-ordered amplitudes.
 However, this relation will introduce complicated kinematic
factors which are functions of the external momenta \cite{Bern:2008qj}.
Although, there has been extensions of the BCJ relation
to cases with fermions e.g., \cite{Sondergaard:2009za,Melia:2013epa},
there is no general expression for the BCJ relation in the
presence of (fundamental representation) quarks.
Since the long term goal of this work is to pursue recursion
for processes involving quarks as well, we will therefore not further
consider BCJ relations.

\item The second approach is to perform the recursion directly in
the multiplet basis.
Once we know the multiplet basis decompositions for
amplitudes for fewer gluons and relations for decomposing the
color structure appearing in the BCFW-recursion,
we can derive a recursion relation for the kinematic factors $A^\alpha$
via the BCFW recursion for the color-dressed amplitudes.
 \end{itemize}

In the following, we will demonstrate the decomposition
using the above strategies. In \secref{sec:comparison}, we calculate
the kinematic factors $A^\alpha$ by evaluating scalar products.
Section~\ref{sec:recursion} provides a derivation of color-dressed
BCFW recursion, followed by a recursion relation for the color
structure of MHV gluon amplitudes formulated in the multiplet basis.
This finally allows us to derive the BCFW recursion for the kinematic factors $A^\alpha$
for any number of gluons.
In \secref{sec:conclusion}, we conclude and discuss natural extensions.

\section{ Evaluation by scalar products}
\label{sec:comparison}
In this section, we derive the multiplet basis expansion by comparing
the color factors in multiplet bases to those in DDM or trace bases.
The general framework for this construction is as follows:
\begin{itemize}
\item  The color vectors (including powers of $\TR$) in the DDM or trace basis (or, in general,
any spanning set in which the recursion is known) can be expanded in
the multiplet basis vectors,
    \bea
     \Col^{\sigma}=\Sl_{\alpha}a^{ \alpha \sigma} \Vec^{\alpha},~~\Label{DDM-Multi}
    \eea
    where the coefficients are given by scalar products of
    these two kinds of color factors\footnote{We assume the orthogonal
      multiplet basis to also be normalized, if it is not,
     this is trivially accounted for.}
   \bea
   a^{\alpha\sigma}  =\langle\Vec^{\alpha}|\Col^{\sigma}\rangle.~~\Label{scalar-product}
   \eea
   The scalar product is given by summing over all external color
   indices, i.e., for arbitrary color structures $\Col^1$ and $\Col^2$,
   \begin{equation}
      \langle\Col^1 | \Col^2\rangle
     =\sum_{a_1,\,a_2,\,...}(\Col^{1}{}_{a_1\,a_2...})^* \, \Col^2{}_{a_1\,a_2...},
     \label{eq:scalar_product}
   \end{equation}
   with $a_i=1,\hdots,\Nc$ if parton $i$ is a quark or antiquark and
   $a_i=1,\hdots,\Nc^2-1$ if parton $i$ is a gluon.
 \item Substituting the expression \eqref{DDM-Multi} into the DDM
   decomposition, \eqref{DDM-decomposition}, or the trace basis
   decomposition, \eqref{Trace-decomposition}, and collecting the
   kinematic factors corresponding to a given basis vector $\Vec^{\alpha}$
   in the multiplet basis, the color-dressed amplitude can be stated as
   \bea
   {\cal M}(g_1,g_2,\dots, g_n)=g^{n-2}\Sl_{\alpha}\left[\Sl_{\sigma}
     a^{\alpha \sigma }
     A(\sigma)\right]\Vec^{\alpha}.~~\Label{multiplet-from-DDM}
   \eea
   Comparing the above expression to the multiplet basis expansion
   \eqref{Multiplet-decomposition1}, we can read off the kinematic
   factor multiplying the basis vector $\Vec^\alpha$
   \bea
    A^{\alpha}=\Sl_{\sigma}
   a^{\alpha \sigma }
   A(\sigma).~~\Label{kinematic-from-DDM}
   \eea
\end{itemize}
In principle, one can use this method to derive the
multiplet basis expansions for an arbitrary number of external legs with arbitrary helicity configuration.
We have calculated the multiplet basis decompositions for up to
six gluons, and demonstrate the calculations in the
three- and four-gluon cases below.
The five- and six-gluon cases are treated using multiplet basis
recursion in \secref{sec:recursion}.
\subsection{The three-gluon amplitude}
For three gluons the multiplet basis can be chosen 
as
\bea
  \Vec_{g_1\,g_2\,g_3}^{8s}&=& d^{g_1 g_3 g_2}=\frac{1}{\TR}[\Tr(t^{g_1}t^{g_3}t^{g_2})+\Tr(t^{g_1}t^{g_2}t^{g_3})],\nonumber \\
  \Vec_{g_1\,g_2\,g_3}^{8a}&=&if^{g_1 g_3 g_2}=\frac{1}{\TR}[\Tr(t^{g_1}t^{g_3}t^{g_2})-\Tr(t^{g_1}t^{g_2}t^{g_3})],
\eea
making it orthogonal but not normalized.
From these two equations, we get
\bea
  \Tr(t^{g_1}t^{g_3}t^{g_2})=\frac{\TR}{2} \left[d^{g_1 g_3 g_2}+if^{g_1 g_3 g_2}\right],
~~\Tr(t^{g_1}t^{g_2}t^{g_3})=\frac{\TR}{2} \left[d^{g_1 g_3 g_2}-if^{g_1 g_3 g_2}\right].
\eea
Thus the three-gluon multiplet basis expansion is given by
\bea
\mathcal{M}({g_1},{g_2},{g_3})&=&g\left(1\over\sqrt{\TR}\right)^3\
\frac{\TR}{2}[d^{g_1 g_3 g_2}+if^{g_1 g_3 g_2}]A(1,3,2)+
g\left(1\over\sqrt{\TR}\right)^3\frac{\TR}{2}[d^{g_1 g_3 g_2}-if^{g_1 g_3 g_2}] A(1,2,3)\nn
&=&
g\frac{1}{2\sqrt{\TR}} d^{g_1 g_3 g_2}[A(1,3,2)+A(1,2,3)]+
g\frac{1}{2\sqrt{\TR}}if^{g_1 g_3 g_2}[A(1,3,2)-A(1,2,3)].
\eea
Since $A(1,2,3)$ is antisymmetric under $1\leftrightarrow 2$ we find
\bea
A^{8s}=0, \quad  A^{8a}={1\over \sqrt{\TR}} A(1,3,2),
\eea
where we note that the first equation can be seen as a
manifestation of charge conjugation invariance (cyclic reflection),
and that the second color factor is precisely the amplitude for the
(only) vector in the DDM basis.

\subsection{The four-gluon amplitude}
In the four-gluon case, we start from the DDM decomposition
with $1$ and $2$ as the fixed legs
\bea
\mathcal{M}(g_1,g_2,g_3,g_4)=
	g^2{1\over\TR}if^{g_1g_2i_1} if^{i_1g_3g_4}A(1,2,3,4)+
	g^2{1\over\TR}if^{g_1g_3i_1} if^{i_1g_2g_4}A(1,3,2,4).
\eea
Using \ColorMath \cite{Sjodahl:2012nk} to evaluate the
scalar products in \eqref{scalar-product} this is decomposed
into the multiplet basis\footnote{The gluon order convention in $\Vec^\alpha_{g_1\,g_3;\,g_2\,g_4}$ probably seems unnatural at this stage, but the advantages will become clear in \secref{sec:color_recursion}.}
$\mathcal{V}$
\bea
\mathcal{V}=\Bigl\{\Vec_{g_1\,g_3;\,g_2\,g_4}^{1}, \Vec_{g_1\,g_3;\,g_2\,g_4}^{8s}, \Vec_{g_1\,g_3;\,g_2\,g_4}^{8a} , \Vec_{g_1\,g_3;\,g_2\,g_4}^{27} ,
\frac{1}{\sqrt{2}}\left[\Vec_{g_1\,g_3;\,g_2\,g_4}^{10}+ \Vec_{g_1\,g_3;\,g_2\,g_4}^{\tenbar}\right], \Vec_{g_1\,g_3;\,g_2\,g_4}^{0}\Bigr\},
\label{eq:4gNormalized}
\eea
given by \cite{Keppeler:2012ih},
\bea
\Vec_{g_1\,g_3;\,g_2\,g_4}^\alpha=\frac{1}{\sqrt{d_\alpha}}\Proj^\alpha_{g_1\,g_3;\,g_2\,g_4}
\eea
where  $\Proj^{\alpha}$ is the projection operator onto the
irreducible representation $\alpha$ with dimension $d_{\alpha}$
\cite{Butera:1979na,Cvi84,Cvi08,Dokshitzer:2005ig,Keppeler:2012ih},
\begin{eqnarray}
  \Proj^{1}_{g_1\, g_3;\,g_2\,g_4}
  &=& \frac{1}{\Nc^2-1}\delta^{{g_1\,g_3}} \delta^{{g_2\,g_4}} \, ,
  \nonumber \\
  \Proj^{8s}_{g_1\, g_3;\,g_2\,g_4}
  &=& \frac{\Nc}{2 \TR (\Nc^2-4)} d^{{g_1 g_3 i_1}} d^{{i_1 g_2 g_4}} \, ,
  \nonumber\\
  \Proj^{8a}_{g_1\, g_3;\,g_2\,g_4}
  &=& \frac{-1}{2 \Nc \TR} i f^{g_1\,g_3\,i_1} i f^{i_1\,g_2\,g_4} \, ,
  \nonumber\\
  \Proj^{27}_{g_1\, g_3;\,g_2\,g_4}
  &=& \frac{1}{4} (\delta^{g_1\,i_1} \delta^{g_3\,i_2}
  + \delta^{g_1\,i_2} \delta^{g_3\,i_1})
  \left[ \delta^{i_1\,g_2}\delta^{i_2\,g_4}
    + \frac{1}{\TR^2} \tr(t^{i_1} t^{g_4} t^{i_2} t^{g_2}) \right]
  \nonumber\\ &&
  - \frac{\Nc-2}{2\Nc} \, \Proj^{8s}_{g_1\, g_3;\,g_2\,g_4}
  - \frac{\Nc-1}{2\Nc} \, \Proj^{1}_{g_1\, g_3;\,g_2\,g_4} \, ,
  \nonumber\\
  \Proj^{10}_{g_1\, g_3;\,g_2\,g_4}
  &=& \frac{1}{4} (\delta^{g_1\,i_1} \delta^{g_3\,i_2}
  - \delta^{g_1\,i_2} \delta^{g_3\,i_1})
  \left[ \delta^{i_1\,g_2}\delta^{i_2\,g_4}
    + \frac{1}{\TR^2} \tr(t^{i_1} t^{g_4} t^{i_2} t^{g_2}) \right]
  - \frac{1}{2} \, \Proj^{8a}_{g_1\, g_3;\,g_2\,g_4} \, ,
  \nonumber\\
  \Proj^{\overline{10}}_{g_1\, g_3;\,g_2\,g_4}
  &=& \frac{1}{4} (\delta^{g_1\,i_1} \delta^{g_3\,i_2}
  - \delta^{g_1\,i_2} \delta^{g_3\,i_1})
  \left[ \delta^{i_1\,g_2}\delta^{i_2\,g_4}
    - \frac{1}{\TR^2} \tr(t^{i_1} t^{g_4} t^{i_2} t^{g_2}) \right]
  - \frac{1}{2} \, \Proj^{8a}_{g_1\, g_3;\,g_2\,g_4} \, ,
  \nonumber\\
  \Proj^{0}_{g_1\, g_3;\,g_2\,g_4}
  &=& \frac{1}{4} (\delta^{g_1\,i_1} \delta^{g_3\,i_2}
                   + \delta^{g_1\,i_2} \delta^{g_3\,i_1})
      \left[ \delta^{i_1\,g_2}\delta^{i_2\,g_4}
             - \frac{1}{\TR^2} \tr(t^{i_1} t^{g_4} t^{i_2} t^{g_2}) \right]
      \nonumber\\ &&
      - \frac{\Nc+2}{2\Nc} \, \Proj^{8s}_{g_1\, g_3;\,g_2\,g_4}
      - \frac{\Nc+1}{2\Nc} \, \Proj^{1}_{g_1\, g_3;\,g_2\,g_4} \, ,
  \label{eq:ggBasis}
\end{eqnarray}

and the general expressions for the dimensions are\footnote{Independently of $\Nc$ we refer to the adjoint
representation as the octet representation, and similarly
we label other representations by their $\Nc=3$ dimension,
although clearly the dimension depends on $\Nc$.}
\bea\label{eq:RepresentationDimensions}
d_{8}=\Nc^2-1,\;\;
d_{10}=\frac{1}{4}(\Nc^4 - 5 \Nc^2 + 4),\;\;
d_{27}=\frac{1}{4}\Nc^2(\Nc^2 + 2 \Nc - 3)\;\;
d_{0}=\frac{1}{4}\Nc^2(\Nc^2 - 2 \Nc - 3).\;\;
\eea
Expressed in this basis (which is also electronically attached as an online
resource) the amplitude can be stated
\bea
\mathcal{M}(g_1,g_2,g_3,g_4)
=g^2\mathcal{A}\cdot\mathcal{V},
\eea
where $\mathcal{A}$ is the kinematic factor,
\bea
\mathcal{A}&=&\Nc 
\times\Bigl\{-2 A(1,2,3,4),-\sqrt{\left(\Nc+1\right)\left(\Nc-1\right)}
A(1,2,3,4),-\sqrt{\left(\Nc+1\right)\left(\Nc-1\right)}\\
&&\left[A(1,2,3,4)+2 A(1,3,2,4)\right], \sqrt{\left(\Nc+3\right)\left(\Nc-1\right)}A(1,2,3,4),0,-
\sqrt{\left(\Nc+1\right)\left(\Nc-3\right)}A(1,2,3,4)\Bigr\}.\nonumber\Label{4pt-kinematic-part}
\eea
Note that in the above discussion, we did not specify the
helicity of the external legs.
When we want to consider the kinematic factor for a particular helicity
configuration, e.g., $1^-,2^+,3^+,4^-$ we just substitute the corresponding
form of the color-ordered amplitudes into \eqref{4pt-kinematic-part}.

A few remarks on the basis are in place.
First we note that for $\Nc=3$ the last basis vector vanishes
as it corresponds to a multiplet which only appears
for $\Nc\ge4$. For four gluons this reduction in
dimension due to small $\Nc$ is rather unimportant, but
for large $\Ng$, the difference becomes significant,
(cf. \tabref{tab:RadiationMatrixNonZero}).

 Then we note that, due to charge conjugation invariance (which at tree-level
manifests itself as cyclic reflection in trace bases), the decuplet
and anti-decuplet in \eqref{eq:4gNormalized} must
multiply the same amplitude, which -- at tree-level --
vanishes. Charge conjugation invariance is also the
reason why the octet vectors corresponding to contracting
one antisymmetric structure constant with one symmetric
structure constant vanish.
This means that the four-gluon basis vectors
can be expressed in terms of projectors {\it only}.
Note, however, this a special feature of the
four-gluon basis, it is {\it not} generally possible
(even for $\Ng$ even), as there are different
ways of building up the same multiplet,
see for example \cite{Keppeler:2012ih}.

Finally we point out that although the multiplet decomposed
result \eqref{4pt-kinematic-part} may look somewhat complicated,
it is now in an excellent form for squaring.
For given helicities and external momenta, the color-ordered amplitudes in
\eqref{4pt-kinematic-part} are just (complex) numbers. To get the full amplitude
square, we thus just have to square the coefficients in \eqref{4pt-kinematic-part}
 and
add them up.

In this particular case of four gluons only, not much is gained
by this rewriting, as the scalar product matrix for the DDM
basis anyway only involves $2\times 2$ terms.
However, for larger bases, where the $2\times2$ scalar product
matrix is replaced by a matrix of dimension $(\Ng-2)!\times(\Ng-2)!$
(or $(\Ng-1)!\times(\Ng-1)!$ for trace bases), avoiding the factorial
square is clearly desirable.

Unfortunately, with the naive way of calculating scalar products
utilized here, involving direct evaluation of
$(\Ng-2)!$ $\times$ (the number of multiplet basis vectors) entries,
what is gained in the squaring step for multiplet bases,
may be lost in the step of scalar product decomposition.
We do note, however, that a more clever procedure for evaluating
scalar products, based on the birdtrack method and Wigner $3j$ and $6j$
coefficients \cite{Cvi08,Sjodahl:2015qoa} possibly could
change this conclusion.
As it is unclear if the scalar product method is beneficial,
the remainder of this paper instead focuses on deriving
recursion relations directly in the multiplet basis.

\section{Recursion in multiplet bases}
\label{sec:recursion}
In this section, we present an on-shell recursion approach for
the kinematic factor ${A}^{\alpha}$ in the multiplet basis expansion
\eqref{Multiplet-decomposition1}. We will show that once we know
the recursion relation between the color factors in the multiplet
bases for the $n$-gluon and $(n-1)$-gluon amplitudes in addition to
the BCFW 
recursion for color-dressed amplitudes, we can derive a recursion
relation for the kinematic factors ${A}^{\alpha}$ for the MHV
helicity configuration. The main idea is:
\begin{itemize}
\item We use BCFW recursion to rewrite the color-dressed amplitude
in terms of products of on-shell lower point color-dressed
amplitudes for all MHV channels.
\item  For a given channel in the BCFW recursion, the color factors in the
BCFW expression of the $\Ng$-gluon amplitude can be
constructed by vectors in the $(\Ng-1)$-gluon multiplet basis
contracted with structure constants,
while the corresponding kinematic factor for the BCFW
expression is obtained from the $(n-1)$-gluon on-shell MHV amplitude
and the three-gluon $\overline{\mbox{MHV}}$ amplitude.
\item We derive a recursion relation for the color structure
between the $\Ng$- and $(\Ng-1)$-gluon multiplet bases.
Using this recursion relation we can express the color
factors in the $\Ng$-gluon multiplet basis, and collecting
the kinematic factors corresponding to the same multiplet basis vector, we
obtain a recursion relation for the $n$-gluon kinematic factor $A^{\alpha}$.
\end{itemize}
In the following, we first present a review of BCFW recursion
for color-dressed amplitudes and then discuss the MHV configuration.

\subsection{BCFW recursion for color-dressed amplitudes}
\begin{figure}
  \centering
  \includegraphics[width=0.7\textwidth]{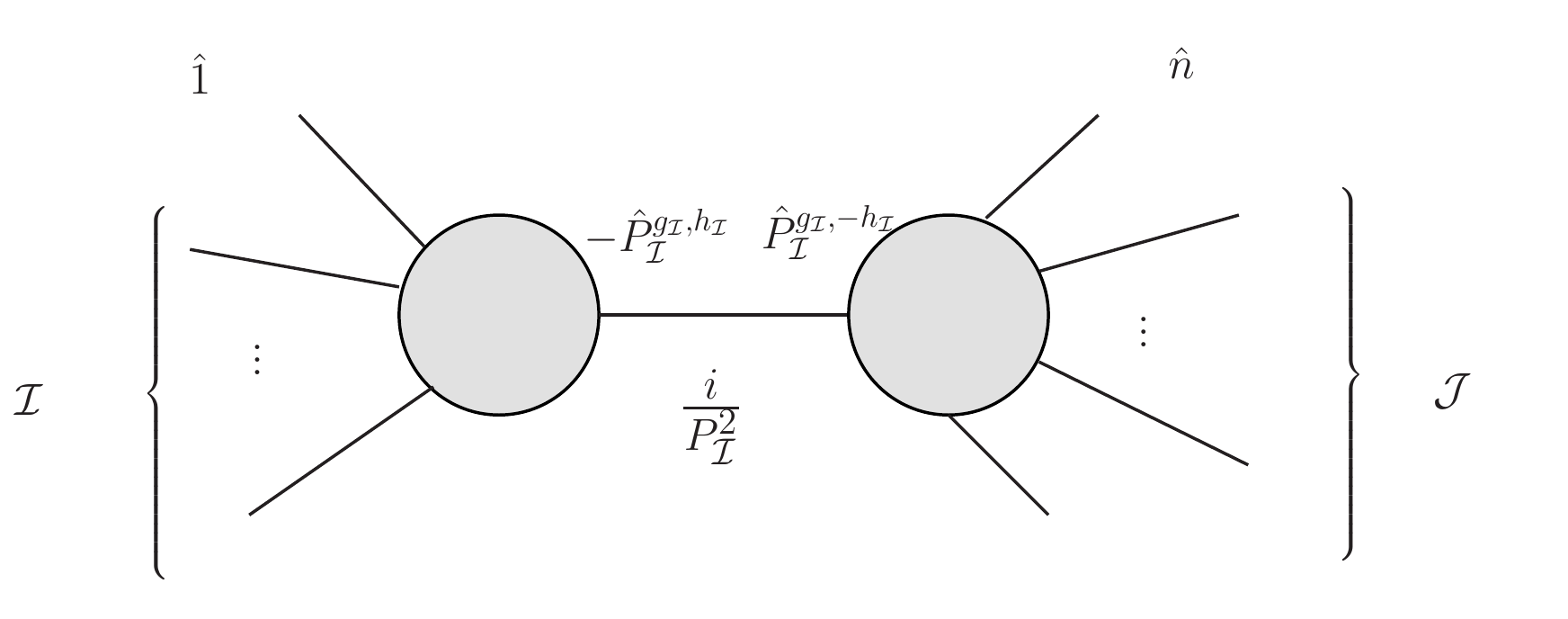}
  \caption{The amplitude is divided into a left set ${\cal I} \cup \hat{1}$
    and a right set ${\cal J}\cup \hat{n}$. Note that here, as opposed to in
\figref{BCFW}, no ordering is inferred among the partons on the left and right side.} \label{fig:BCFW2}
\end{figure}

Now let us review the BCFW recursion for color-dressed gluon
amplitudes at tree-level  \cite{Duhr:2006iq}.
We consider an $n$-gluon color-dressed tree amplitude
$\mathcal{M}({g_1},{g_2},\dots,{g_n})$, where $g_i$ is used to denote
a gluon with momentum (counted outwards), helicity and color.
If we shift the momenta of the gluons $g_1$ and $g_n$
with a complex four-vector $q$ obeying
\bea
q\cdot p_1=q\cdot p_n=q^2=0,
\label{eq:on-shell}
\eea
the shifted momenta,
\bea
\WH p_1(z)=p_1-zq,~~\WH p_n(z)=p_n+zq,~~\Label{Shift}
\eea
 remain on-shell.
With this shift, the color-dressed amplitude $\mathcal{M}(z)$
becomes a rational function of the complex variable $z$.
The desired amplitude is just $\mathcal{M}(0)$. To solve for
$\mathcal{M}(0)$, we  use Cauchy's theorem
\bea
\oint\limits_{\text{finite poles}} dz {\mathcal{M}(z)\over z}
=\oint\limits_{z\rightarrow\infty}dz{\mathcal{M}(z)\over z}.
\eea
The integrals around the finite poles are given by their residues, thus we have
\bea
\mathcal{M}(0)=-\Sl_{z_i\neq 0}\text{Res}_{z\rightarrow z_i}{\mathcal{M}(z)\over z}+\mathcal{B},
\eea
where $\mathcal{B}$ comes from the contour integral at infinity.
 In the study of BCFW recursion, the boundary behavior when
$z\rightarrow \infty$ is important and has been investigated systematically
in \cite{ArkaniHamed:2008yf}.
For gluon amplitudes, we can always choose a shift such that
${\cal B}=0$. The residues of the finite poles can be obtained
by considering the factorization behavior, with which the amplitude
is factorized into two on-shell sub-amplitudes when an internal
line goes on-shell. The nontrivial contributions for the $z$-poles
are those with the two shifted legs in two different
sub-amplitudes. For the shift in \eqref{Shift}, we let  the gluon
$g_1$ be in the left sub-amplitude and the gluon $g_n$ be in the right
sub-amplitude. If we divide the other $(n-2)$  gluons into the left
set ${\cal I}$ and the right set ${\cal J}$, as in \figref{fig:BCFW2} ,
the position of the pole corresponding to this division can be found by solving

\bea
\WH P_{\cal I}^2(z)=\left(\WH p_1(z)+\Sl_{k\in {\cal I}}p_k\right)^2=0,
\eea
giving the poles
\bea
z_{\cal I}={\left(p_1+\Sl_{k\in {\cal I}}p_k\right)^2\over 2q\cdot \left(p_1+\Sl_{k\in {\cal I}}p_k\right)}.~~\Label{pole}
\eea
The $n$-gluon color-dressed amplitude is factorized into two lower point on-shell amplitudes
\bea
\mathcal{M}(z)\stackrel{z\to z_{\cal I}}
{\longrightarrow}
\Sl_{ i_{\cal I},h_{\cal I}}
\mathcal{M}\left(\WH{g}_1( z_{\cal I}),\{{g_k}\mid_{k\in{\cal I}}\},-\WH P^{ i_{\cal I},h_{\cal I}}_{\cal I}(z_{\cal I})\right){ i\over \WH P^2_{\cal I}(z_{\cal I})}\mathcal{M}\left(\WH P^{i_{\cal I},-h_{\cal I}}_{\cal I}(z_{\cal I}),\{{g_l}\mid_{l\in\cal J}\},\WH{g}_n(z_{\cal I})\right),
\eea
where $ i_{\cal I}$ and $h_{\cal I}$ are used to denote the color and helicity indices for the internal line.
Then at $z_{\cal I}$
\bea
{\mathcal{M}(z)\over z}{ \stackrel{z\to z_{\cal I}}{\longrightarrow}}\Sl_{{ i_{\cal I}},h_{\cal I}} \mathcal{M}\left(\WH {g}_1(z_{\cal I}),\{{g_k}\mid_{k\in{\cal I}}\},-\WH P^{i_{\cal I},h_{\cal I}}_{\cal I}(z_{\cal I})\right){i\over z_{\cal I}\WH P^2_{\cal I}(z_{\cal I})}\mathcal{M}\left(\WH P^{ i_{\cal I},-h_{\cal I}}_{\cal I}(z_{\cal I}),\{{g_l}\mid_{l\in{\cal J}}\},\WH{g}_n(z_{\cal I})\right).
\eea
Considering the position of the pole in \eqref{pole}, we have
\bea
{\mathcal{M}(z)\over z}\stackrel{z\to z_{\cal I}}{\longrightarrow}-\Sl_{ i_{\cal I},h_{\cal I}}{i{ \mathcal{M}\left(\WH {g}_1(z_{\cal I}),\{{g_k}\mid_{k\in{\cal I}}\},-\WH P_{\cal I}^{ i_{\cal I},h_{\cal I}}(z_{\cal I})\right)\mathcal{M}\left(\WH P^{ i_{\cal I},-h_{\cal I}}_{\cal I}(z_{\cal I}),\{{g_l}\mid_{l\in{\cal J}}\},\WH {g}_n(z_{\cal I})\right)\over P^2_{\cal I}}}{1\over z-z_{\cal I}},
\eea
where the denominator is the squared sum of the momenta of the
external legs in the left set.
Thus the residue for the division ${\cal I}, {\cal J}$ is
\bea
\operatornamewithlimits{\text{Res}}_{z\rightarrow z_{\cal I}}{\mathcal{M}(z)\over z}
&=&
-\Sl_{ i_{\cal I},h_{\cal I}}{ \mathcal{M}\left(\WH {g}_1(z_{\cal I}),\{{g_k}\mid_{k\in{\cal I}}\},-\WH P^{ i_{\cal I},h_{\cal I}}_{\cal I}(z_{\cal I})\right)}{i\over P^2_{\cal I}}{\mathcal{M}\left(\WH P^{ i_{\cal I},-h_{\cal I}}_{\cal I}(z_{\cal I}),\{{g_l}\mid_{l\in{\cal J}}\},\WH {g}_n(z_{\cal I})\right)}.\nn
\eea
The numerator here is a product of two on-shell sub-amplitudes
with the momentum of gluon $g_1$ and $g_n$ shifted. To get the
full amplitude, we should sum over all possible residues at
all finite poles. The amplitude is then given by the following
BCFW recursion relation
\bea
&&\mathcal{M}\left({g_1},{g_2},\dots,{g_n}\right)~~~\Label{Color-dressed-BCFW}\nn
&=&\Sl_{\cal I}\Sl_{ i_{\cal I},h_{\cal I}}{ \mathcal{M}\left(\WH {g}_1(z_{\cal I}),\{{g_k}\mid_{k\in{\cal I}}\},-\WH P^{i_{\cal I},h_{\cal I}}_{\cal I}(z_{\cal I})\right)}{i\over P^2_{\cal I}}{\mathcal{M}\left(\WH P^{ i_{\cal I},-h_{\cal I}}_{\cal I}(z_{\cal I}),\{{g_l}\mid_{l\in{\cal J}}\},\WH g_n(z_{\cal I})\right)}.
\eea

\subsection{Kinematic recursion}
\label{sec:MHV_recursion}
When we calculate an $(n>3)$-gluon color-dressed amplitude for
a given helicity configuration, the configurations with all
 helicities positive and all helicities except one positive
have to vanish \cite{Berends:1987me}. The first nontrivial
configuration is the MHV configuration with two negative
helicity gluons.

It is convenient to use the spinor helicity formalism
\cite{Berends:1981rb, Causmaecker:1981bg, Kleiss:1985yh, Xu:1986xb, Gunion:1985vca}
to study amplitudes.  In the spinor helicity formalism,
one expresses external momenta $p^{\mu}_i$ by double spinors
$(\la_i)_{a}(\W\la_i)_{\dot{a}}$, where $(\la_i)_{a}$ and $(\W\la_i)_{\dot{a}}$
are two-dimensional Weyl spinors. The polarization vectors
are explicitly expressed as
$\varepsilon^{\mu+}\sim-\sqrt{2}{{\mu_{a}\W\lambda_{\dot{a}}}\over \Spaa{\mu\mid\lambda}}$ 
and
$\varepsilon^{\mu -}\sim-\sqrt{2}{{\lambda_{a}\W\mu_{\dot{a}}}\over \Spbb{\W\lambda\mid\W\mu}}$,
where the spinor products are defined by
$\Spaa{\lambda\mid\mu}\equiv\eps^{ba}\lambda_{a}\mu_{b}$ and
$\Spbb{\W\lambda\mid\W\mu}\equiv\eps^{\dot{a}\dot{b}}\W\lambda_{\dot a}\W\mu_{\dot b}$,
using $\mu$ and $\W\mu$ to denote the reference spinors.  
The matrices $\epsilon^{ab}$
and $\epsilon^{\dot a\dot b}$ are given by
\bea
\epsilon^{ab}=\epsilon^{\dot a\dot b}=\left(
\begin{array}{cc}
  0 & 1 \\
  -1 & 0 \\
\end{array}
\right).
\eea
In the spinor helicity formalism, all amplitudes are expressed by spinor products. The color-ordered MHV amplitude is given by the famous formula
\bea
A\left(1^+,2^+,\dots,i^-,\dots,j^-,\dots,n^+\right)=i{\Spaa{i\mid j}^4\over \Spaa{1\mid 2}\Spaa{2\mid 3}\dots\Spaa{{n-1}\mid n}\Spaa{n\mid 1}},~~\Label{MHV-amplitude}
\eea
which was conjectured in \cite{Parke:1986gb} and proven in \cite{Berends:1987me}.
When we consider the $\overline{\text{MHV}}$ amplitude where all helicities are flipped,
we just (up to a factor $(-1)^n$) replace $\Spaa{{}\mid{}}$ by $\Spbb{{}\mid{}}$
\bea
A\left(1^-,2^-,\dots,i^+,\dots,j^+,\dots,n^-\right)=(-1)^ni{\Spbb{i\mid j}^4\over \Spbb{1\mid 2}\Spbb{2\mid 3}\dots\Spbb{{n-1}\mid n}\Spbb{n\mid 1}}.
\eea

Let us now consider the color-dressed MHV amplitude with
$g_1$ and $g_n$ as negative helicity gluons. We can
conveniently shift the momenta of $g_1$ and $g_n$
in the spinor helicity formalism
\bea
& &\lambda_1\rightarrow\lambda_1,\;\;\; \W\lambda_1\rightarrow\W\lambda_1-z\W\lambda_n,\nn
& &\W\lambda_n\rightarrow \W\lambda_n,\;\;\; \lambda_n\rightarrow \lambda_n+z\lambda_1.
\label{eq:spinor_shift}
\eea
This is nothing but the spinor expression of the $(1,n)$-shift
defined by \eqref{Shift}, as $p_1\sim \lambda_1\W \lambda_1$, $p_n\sim\lambda_n\W\la_n$,
and $q\sim\la_1\W\la_n$. The constraint equations, \eqsref{eq:on-shell},
forcing the momenta in \eqref{Shift} to remain on-shell,
are automatically satisfied in this spinor expression and the boundary
contribution $\mathcal{B}$ vanishes, meaning that we can use the
BCFW recursion for the color-dressed amplitude \eqref{Color-dressed-BCFW}
without any boundary correction.
\begin{figure}
  \centering
  \includegraphics[width=0.45\textwidth]{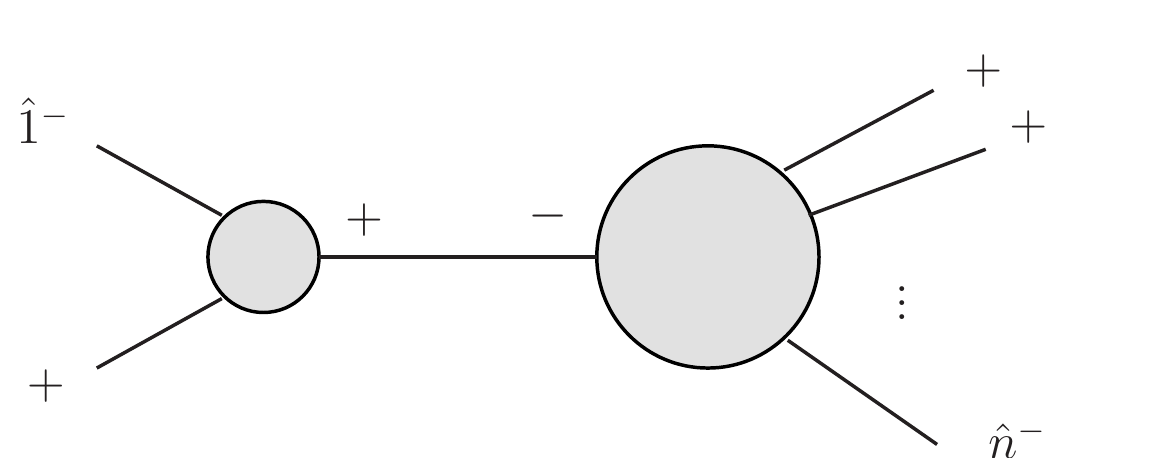}
  \includegraphics[width=0.45\textwidth]{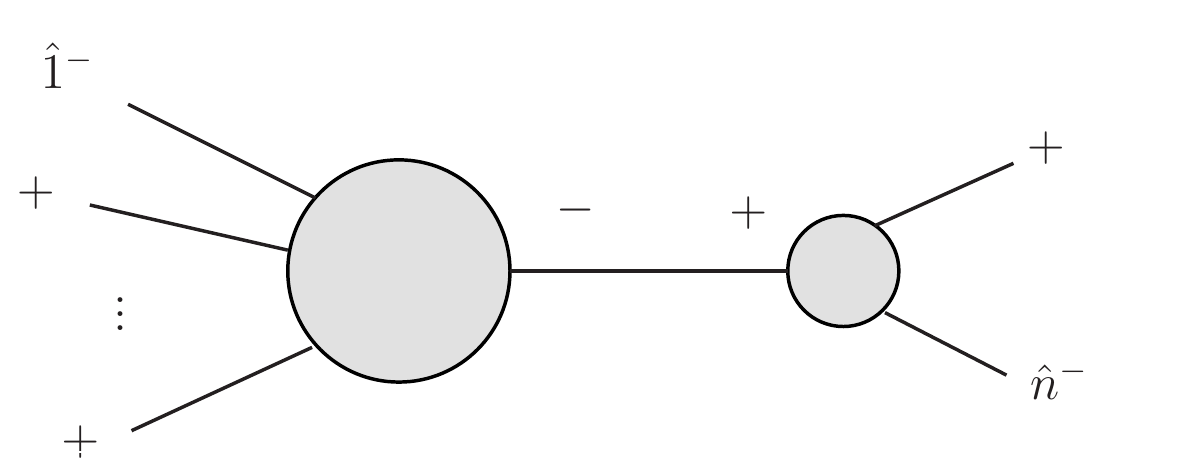}
  \caption{
    The allowed divisions, $(3,n-1)$ (left) and $(n-1,3)$ (right),
    for the MHV amplitude with the gluons with shifted momenta
    $\hat{1}$ and $\hat{n}$ as the negative helicity legs.
    As is proven in \appref{sec:3_n-1division}, only the $(n-1,3)$ factorization channel
    contributes.
  } \label{fig:divisions}
\end{figure}
We also note that with this shift, $g_1$ and $g_n$ must be in
opposite sub-amplitudes.
For MHV amplitudes, there are only two types of divisions,
sketched in \figref{fig:divisions}, which possibly could contribute
\begin{itemize}
\item The $(3,n-1)$ divisions with $3$-gluon $\overline{\text{MHV}}$
amplitudes as left sub-amplitudes and $(n-1)$-gluon MHV amplitudes
as right sub-amplitudes.
\item  The $(n-1,3)$ divisions with $(n-1)$-gluon MHV amplitudes as
left sub-amplitudes and $3$-gluon $\overline{\text{MHV}}$ amplitudes as
right sub-amplitudes.
\end{itemize}
The other divisions always contain sub-amplitudes (for more than three gluons)
with less than two negative helicity gluons and have to vanish.
 In fact, as is proven in \appref{sec:3_n-1division}, the
$(3,n-1)$ division also vanishes.
Thus the full color-dressed amplitude can be stated
\bea
&&\mathcal{M}\left(g_1^{-},g_2^{+},\cdots,g_{n-1}^{+},g_n^{-}\right)~~~~\Label{Color-dressed-BCFW-MHV}\nn
&=&\Sl_{i=2}^{n-1}\mathcal{M}\left(\WH g_1^{-},\dots,g_{i-1}^{+},\WH P^{i_i,-}_{i,n},g_{i+1}^{+},\dots,g^+_{n-1}\right){i\over P^2_{i,n}}\mathcal{M}\left(-\WH P^{ i_i,+}_{i,n},g_i^{+},\WH g_n^{-}\right),
\eea
where $i_i$ denotes the (implicitly summed over) color index
connecting the sub-amplitudes. The right sub-amplitude is given by
\bea
\mathcal{M}\left(-\WH P^{ i_i,+}_{i,n},g_i^{+},\WH g_n^{-}\right)=g{1\over\sqrt{\TR}}if^{g_i i_ig_n}A^{\overline{\text{MHV}}}\left(i^{+},-\WH P^{+}_{i,n},\WH n^{-}\right)=g{1\over\sqrt{\TR}}if^{g_i i_ig_n}(-i){\Spbb{i\mid -\WH P_{i,n}}^3\over\Spbb{-\WH P_{i,n}\mid \WH n}\Spbb{\WH n\mid i} }.\;\;\;
\eea
The left sub-amplitude is given by the $(n-1)$-gluon multiplet basis expansion
\bea
&&\mathcal{M}\left(\WH g_1^{-},\dots,g_{i-1}^{+},\WH P^{ i_i,-}_{i,n},g_{i+1}^{+},\dots,g_{n-1}^+\right)~~~\Label{(n-1)-sub-amplitude}\nn
&=&g^{n-3}\Sl_{\alpha}\Vec^{\alpha}_{g_1 \dots g_i\dots g_{n-1}}|_{g_i\to i_i}{A}^{\alpha}\left(\WH 1^-,\dots,(i-1)^+,\WH P^-_{i,n},(i+1)^+,\dots, (n-1)^+\right),
\eea
 where, at this point, we make no statement about what gluons are
counted as incoming and outgoing in the multiplet bases.
The recursion relation between the $n$-gluon color factors
and the $(n-1)$-gluon color factors can be written as
\bea
\left( \Vec^{\alpha}_{g_1 \dots g_i \dots g_{n-1}}|_{g_i\to  i_i}\right)if^{g_ii_ig_n}&=&
\Sl_{\beta}
 \left(\mathbf{T}_{i}\right)_{\beta \alpha}
\Vec^{\beta}_{g_1 \dots g_n}.\Label{recursion-color-1}
\eea
The matrices $\mathbf{T}_{i}$ describe the effect of emitting
gluon $\Ng$ from gluon $i$ (from the vector $\alpha$ in the $(\Ng-1)$-gluon basis),
decomposed into the $\Ng$-gluon basis. We will refer to these
matrices as the radiation matrices for $(\Ng-1) \rightarrow \Ng$ gluons,
and in \secref{sec:color_recursion}, we will show how to calculate
them efficiently.
Inserting \eqref{(n-1)-sub-amplitude} and \eqref{recursion-color-1} into the BCFW expression for the color-dressed amplitude, \eqref{Color-dressed-BCFW-MHV}, and collecting the kinematic factor corresponding to $\Vec^{\beta}_{g_1 \dots g_n}$, we obtain the recursion relation for the kinematic factor in the MHV configuration
 \bea
 &&{A}^{\beta}\left(1^-,2^+,3^+,\dots, n^-\right)\\
 &=&\Sl_{i=2}^{n-1}\Sl_{\alpha}
 \left(\mathbf{T}_{i}\right)_{\beta \alpha}
\times \left[{A}^{\alpha}\left(\WH 1^-,\dots,(i-1)^+,\WH P^-_{i,n},(i+1)^+,\dots, (n-1)^+\right){i\over s_{i,n}}(-i){1\over\sqrt{\TR}}{\Spbb{i\mid -\WH P_{i,n}}^3\over\Spbb{-\WH P_{i,n}\mid \WH n}\Spbb{\WH n\mid i} }\right],\nonumber
\Label{kinematic-recursion-MHV-0}
\eea
where $s_{i,n}=(p_i+p_n)^2=2p_i\cdot p_n$.

The above recursion relation for the kinematic factors expresses the
kinematic factor of the $n$-gluon MHV amplitude in terms of the $(n-1)$-gluon
MHV amplitude and the three-gluon $\overline{\text{MHV}}$ amplitude.
To calculate the $n$-gluon kinematic factor for the MHV
configuration using \eqref{kinematic-recursion-MHV-0}, we thus use
the kinematic factors in the multiplet basis expansion of the $(n-1)$-gluon
MHV amplitude and the matrices ${\mathbf{T}_{i}}$ (which will be derived in the next section) as input.

\subsection{Color structure recursion}
\label{sec:color_recursion}

Before stating expressions in the multiplet bases we need
to fix our conventions.
The vector space of interest is the overall singlet space
for the involved (incoming plus outgoing) partons.
Clearly the basis vectors for this space can be chosen in many
different ways.
The prescription detailed in \cite{Keppeler:2012ih} constructs
vectors by first constructing gluon projection operators
projecting on irreducible representations for
$\lfloor \Ng/2 \rfloor \to \lfloor \Ng/2 \rfloor  $ gluons.
Following this, basis vectors for processes with up to $\Ng$
gluons can be constructed. (The extension to processes involving quarks
is achieved by grouping the quarks and antiquarks to $\qqbar$-pairs
and noting that each pair transforms either as a singlet or as an
octet.)
For a process with $\Ng$ gluons, the gluons are divided into
$\lceil \Ng/2 \rceil$ ``incoming'' gluons and
$\lfloor \Ng/2 \rfloor$ ``outgoing'' gluons, such that there
are either equally many outgoing and incoming gluons or one more incoming gluon.
For the full set of gluons to transform under a singlet, the overall
representation under which the ``incoming'' gluons transform must match
the overall representation under which the ``outgoing'' gluons transform.
The total dimension of the vector space is thus given by the number of
ways of combining matching ``incoming'' and outgoing representations.

The gluons on either side are then subgrouped such that the first two gluons
transform under representation $\alpha_1$, the first three gluons
transform under representation $\alpha_2$, etc.
The set of representations is collectively referred to as $\alpha$,
and the ``incoming'' gluons are taken to be
$g_1,g_3,...,g_{2\lceil \frac{\Ng}{2} \rceil-1}$,
whereas the ``outgoing'' are given even numbers
$g_2,g_4,...,g_{2\lfloor \frac{\Ng}{2} \rfloor}$.
Using these conventions, and letting single lines
denote the adjoint representation and double lines denote arbitrary representations,
the orthonormal basis vectors are
\bea\label{eq:MultipletBasisVector}
 \Vec^{\alpha}_{g_1\,g_3\,\dots\,g_{2\lceil{n\over 2}\rceil-1};\,g_2\,g_4\,\dots\,g_{2\lfloor{n\over 2}\rfloor}}
&=&
N^{\alpha_1\dots\alpha_{n-3}}\\
& \times&
\raisebox{-0.4\height}{
	\includegraphics[width=0.7\textwidth]{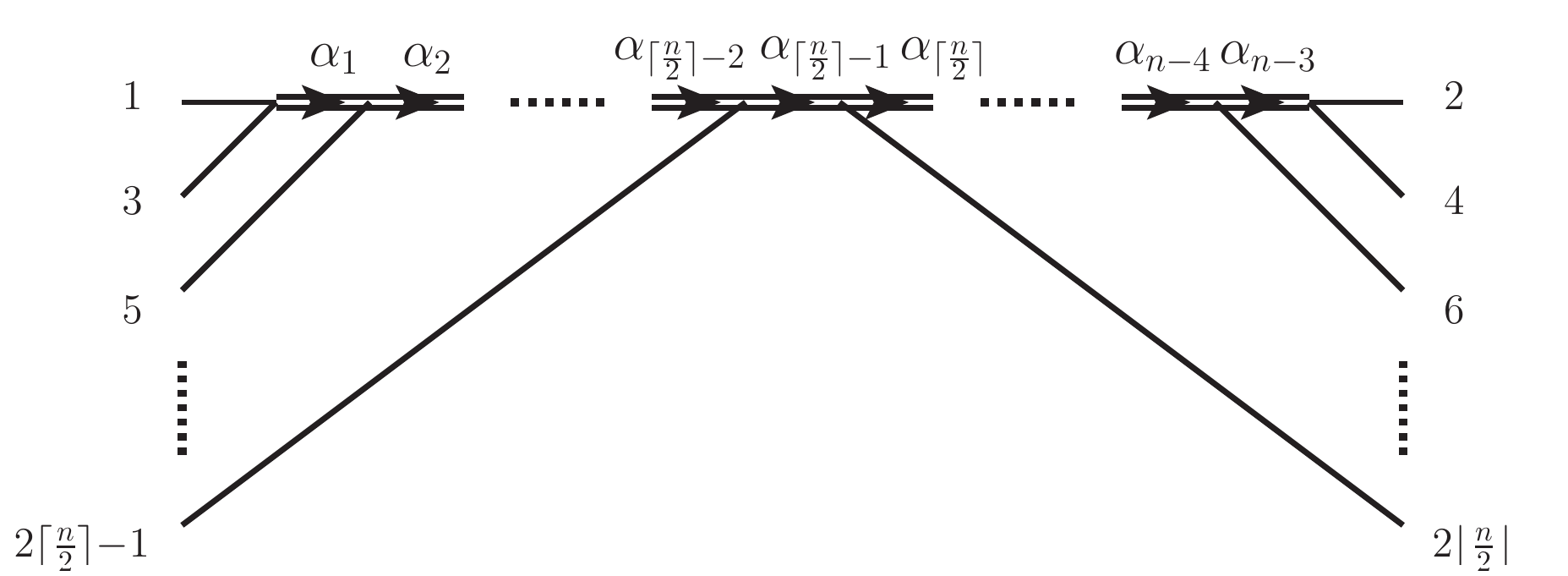}
}\nonumber
\hspace{-4mm},
\eea
where
\begin{equation}\label{eq:MultipletBasisVectorNormalizationFactor}
N^{\alpha_1 \, \alpha_2\dots\alpha_{\Ng-3}}
=
\sqrt{
\frac{
	\prod_{i=1}^{\Ng-3}{d_{\alpha_i}}
}{
	\raisebox{-0.50\height}{\includegraphics[scale=0.3]{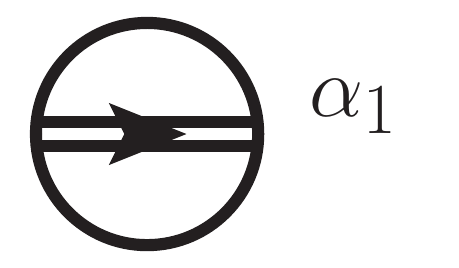}}
	\raisebox{-0.4\height}{\includegraphics[scale=0.3]{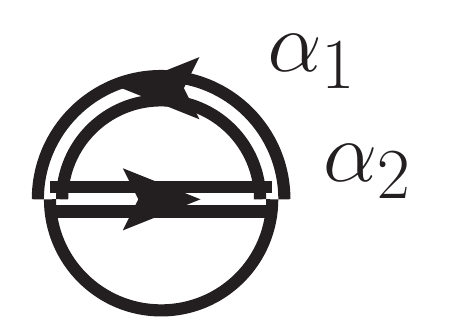}}
	\dots
	\raisebox{-0.43\height}{\includegraphics[scale=0.3]{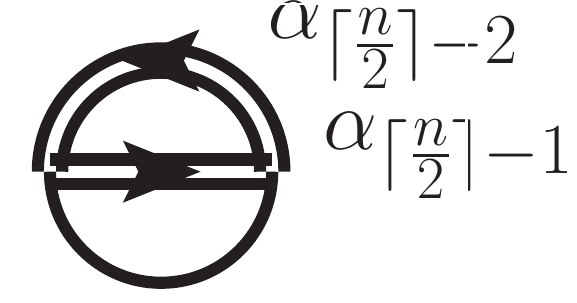}}
	\raisebox{-0.43\height}{\includegraphics[scale=0.3]{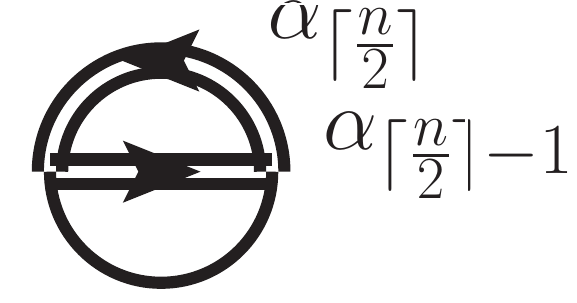}}
	\dots
	\raisebox{-0.4\height}{\includegraphics[scale=0.3]{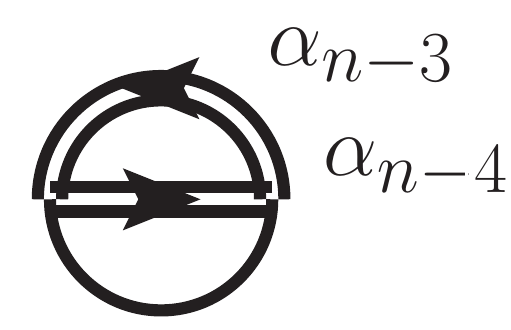}}
	\raisebox{-0.5\height}{\includegraphics[scale=0.3]{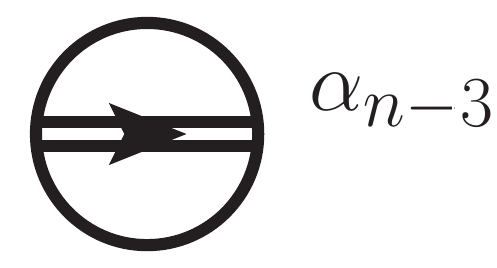}}
}
}.
\end{equation}
Here the vacuum bubbles in the denominator are Wigner $3j$ coefficients.
They can be normalized to one, inducing a normalization for the
generalized vertices connecting the representations.

Letting $A^{\alpha}(1,2,\dots,n)$ denote the amplitude
(for convenience we keep the gluon arguments in $A^{\alpha}$
in this order) the decomposition into these bases may thus be written
\bea
{\cal M}(g_1,g_2,\dots,g_n)=g^{n-2}\Sl_{\alpha}\Vec^{\alpha}_{g_1\, g_3\, ...g_{2\lceil \frac{\Ng}{2} \rceil-1};\, g_2\, g_4\, ...g_{2\lfloor \frac{\Ng}{2} \rfloor}}
A^{\alpha}(1,2,\dots,n).~~\Label{Multiplet-decomposition}
\eea
A remark on the implication of charge conjugation is in place.
As gluons transform under the charge conjugation invariant
adjoint representation, any overall gluon amplitude must respect
this symmetry. This is manifest in the DDM bases in the
sense that each spanning color structure obeys this symmetry,
but it is not manifest in the trace bases and the multiplet bases.
For tree-level trace bases charge conjugation invariance instead shows
up as cyclic reflection. For multiplet bases, charge
conjugation invariance displays itself by the amplitude
for a (non-invariant) basis vector and its conjugate
being equal up to a sign -- and by the vanishing of amplitudes for
which all involved representations are invariant, but the
invariance is spoiled by the generalized vertices.

For many gluons almost all of the basis vectors contain
at least one representation which is not charge conjugation
invariant, meaning that almost every basis vector must occur
with its conjugate. Using these linear combinations
as basis vectors thus reduces the dimension of the
vector space by approximately a factor two.

For the explicit calculations for four, five and six gluons,
we have used conjugation invariant bases. However, for
comparison, the dimensions of the vector spaces,
are -- as for the trace basis case -- stated without
this symmetry in \tabref{tab:RadiationMatrixNonZero}.
This also has the advantage that the vector space dimension for
$\Ng$ gluons is approximately equal to the dimension for
$\Ng-\Nq$ gluons and $\Nq$ $\qqbar$-pairs, see \cite{Sjodahl:2015qoa}.

With the above basis conventions the radiation matrices,
\eqref{recursion-color-1}, are given by
\bea\label{eq:RadiationMatrixDefinition}
\left(\Vec^{\alpha}_{g_1\,g_3\,\dots\,g_{2\lceil{n-1\over 2}\rceil-1};\,g_2\,g_4\,\dots\,g_{2\lfloor{n-1\over 2}\rfloor}}\Big |_{g_i\rightarrow i_i}\right) if^{g_ii_ig_n}
=\sum_{\beta}{
 \left(\mathbf{T}_{i}\right)_{\beta \alpha}
\Vec^{\beta}_{g_1\,g_3\,\dots\,g_{2\lceil{n\over 2}\rceil-1};\,g_2\,g_4\,\dots\,g_{2\lfloor{n\over 2}\rfloor}}
},
\eea
giving for the amplitudes, \eqref{kinematic-recursion-MHV-0},
\bea
  &&  {A}^{\beta}\left(1^-,2^+,3^+,\dots, n^-\right)=
    \Sl_{i=2}^{n-1}\Sl_{\alpha}
 \left(\mathbf{T}_{i}\right)_{\beta \alpha}
\Label{kinematic-recursion-MHV}\nn
&&  \times \left[{A}^{\alpha}\left(\WH 1^-,2^+,\dots, (i-1)^+,\WH P^-_{i,n},(i+1)^+,\dots ,(n-1)^+\right){ 1\over s_{i,n}}
  {1\over\sqrt{\TR}}{\Spbb{i\mid -\WH P_{i,n}}^3\over\Spbb{-\WH P_{i,n}\mid \WH n}\Spbb{\WH n\mid i} }\right].
\eea
As can be seen in \eqref{eq:RadiationMatrixDefinition},
the color structure of the recursion relation in the multiplet bases
is given by inserting one gluon to the $(n-1)$-gluon basis vectors,
\begin{equation}
\left( \Vec^{\alpha}_{g_1\,g_3\,\dots\,g_{2\lceil{n-1\over 2}\rceil-1};\,g_2\,g_4\,\dots\,g_{2\lfloor{n-1\over 2}\rfloor}}
\Big |_{g_i\rightarrow i_i}\right) if^{g_i i_ig_n}.
\label{eq:radiation_color}
\end{equation}
For example, denoting the five-gluon basis vectors $\Vec^{\alpha}_{g_1\,g_3\,g_5;\,g_2\,g_4}$,
if we radiate another gluon $g_6$, we can attach it to any of the gluons
$g_1$, $g_2$, $g_3$, $g_4$ and $g_5$. The corresponding color factors then become
$\Vec^{\alpha}_{i\,g_3\,g_5;\,g_2\,g_4}if^{g_1 i g_6}$,
$\Vec^{\alpha}_{g_1\,g_3\,g_5;\,i\,g_4}if^{g_2  i g_6}$,
$\Vec^{\alpha}_{g_1\,i\,g_5;\,g_2\,g_4}if^{g_3 i g_6}$,
$\Vec^{\alpha}_{g_1\,g_3\,g_5;\,g_2\,i}if^{g_4 i g_6}$ and
$\Vec^{\alpha}_{g_1\,g_3\,i;\,g_2\,g_4}if^{g_5 i g_6}$.
The systematic evaluation of such color structure in the larger
basis is the topic of the present section.

One way of evaluating the radiation matrices is to simply
calculate scalar products between the left hand side color structure
of \eqref{eq:RadiationMatrixDefinition} and the vectors in
the larger basis. This is equivalent to the method of \secref{sec:comparison}.
However, as most of these scalar products
vanish -- for reasons that will become clear later in
this section -- such a strategy would be unnecessarily expensive.

Instead we here present a more elegant way of
evaluating the weights for the $n$-gluon basis
vectors using group theory and the birdtrack notation \cite{Cvi08}.
By applying group theoretical relations to the left hand
side of \eqref{eq:RadiationMatrixDefinition} it can be cast
into the form of the right hand side,
with the radiation matrix elements expressed in terms of
group theoretical weights, the Wigner coefficients.

A method for evaluating scalar products between Feynman diagrams
and multiplet basis vectors is explored in
\cite{Sjodahl:2015qoa}. The same techniques are applicable for
the decomposition of an $(n-1)$-gluon multiplet basis vector which has radiated
an $n$th gluon, into $n$-gluon multiplet basis vectors.
For this decomposition, three group theoretical
relations are required, the completeness relation for tensor products,
the color structure of a vertex correction and the relation between
the ordering of the representations of a vertex.
In birdtrack notation the completeness relation reads
\begin{eqnarray}\label{eq:CompletenessRelation}
  \raisebox{-0.325\height}{
    \includegraphics[scale=0.45]{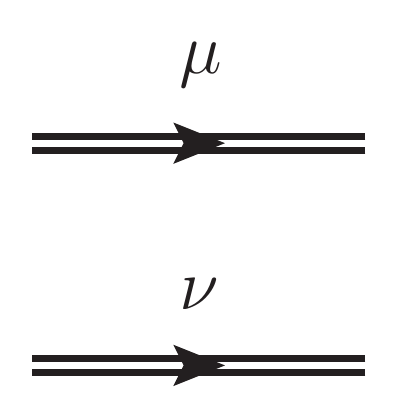}
  }
  =
    \sum_{\alpha\in{}\mu\otimes\nu}{
      \frac{d_\alpha}{\includegraphics[scale=0.3]{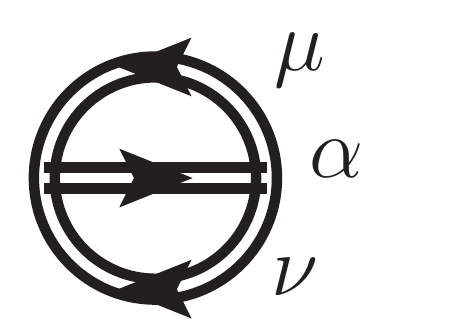}}
      \raisebox{-0.325\height}{
        \includegraphics[scale=0.45]{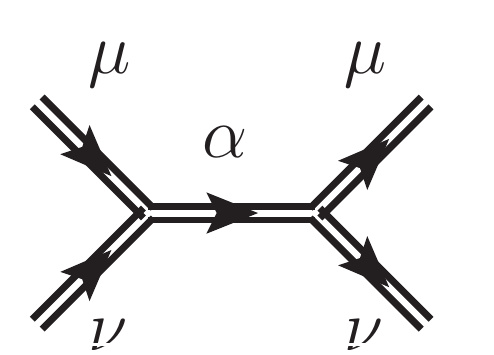}
      }
    }
    \hspace{-1mm}.
\end{eqnarray}
In the tensor product $\mu\otimes\nu$ above, there can be
more than one instance of a particular representation.
In this case all instances have to be summed over,
for example in $\Adj \otimes \Adj $, where $\Adj$ denotes the adjoint representation (not to be confused with the amplitude), there are two ``octets''.

The second relation is a special case of the Wigner-Eckart theorem,
the color structure of a vertex correction can be written as \cite{Sjodahl:2015qoa,Cvi08}
\begin{equation}\label{eq:ColorStructureVertexCorrection}
\raisebox{-0.44\height}{
	\includegraphics[scale=0.45]{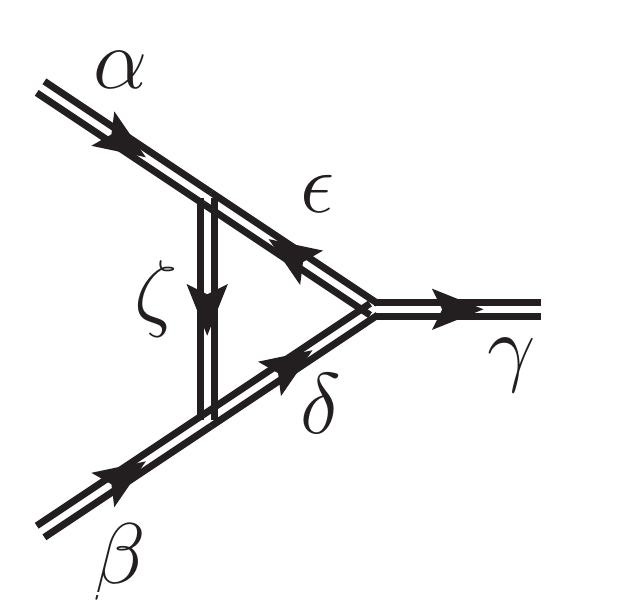}
}
\hspace{-2mm}
=
\sum_{ a 
}{
\frac{
\raisebox{-0.45\height}{
	\includegraphics[scale=0.45]{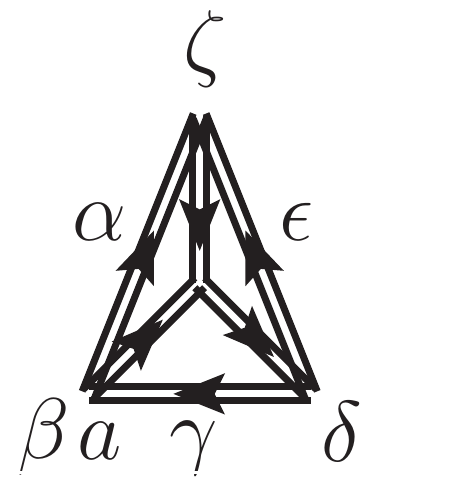}
}
\hspace{-3mm}
}{
\raisebox{-0.45\height}{
	\includegraphics[scale=0.3]{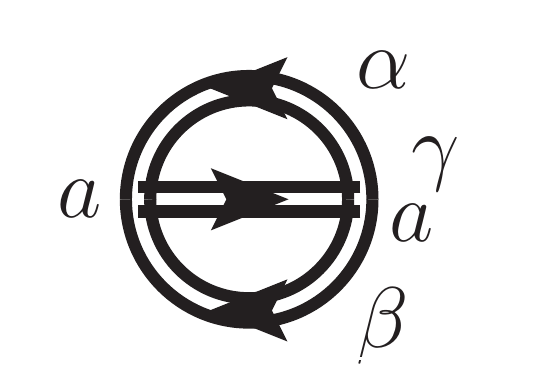}
}
\hspace{-3mm}
}
\raisebox{-0.43\height}{
	\includegraphics[scale=0.45]{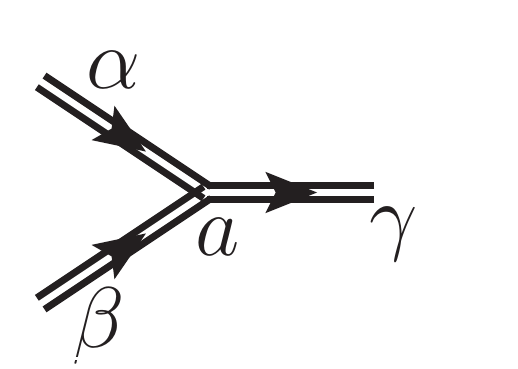}
}
}
\hspace{-3mm}.
\end{equation}
The above equation is sometimes stated without the sum.
Indeed the sum is only needed if there is more than one
instance of $\gamma$ in the tensor product $\alpha\otimes\beta$.
In this case, the $\alpha\beta\gamma$-vertex may appear
in more than one version, this occurs, for example, when $\alpha$, $\beta$ and $\gamma$ are octets, then $a$ is $if$ and $d$.

The third relation concerns
the ordering of representations in a vertex, the relation between
the two orderings is given by \cite{Cvi08},
\begin{equation}\label{eq:YutsisVertexToNormalVertex}
\raisebox{-0.41\height}{
	\includegraphics[scale=0.45]{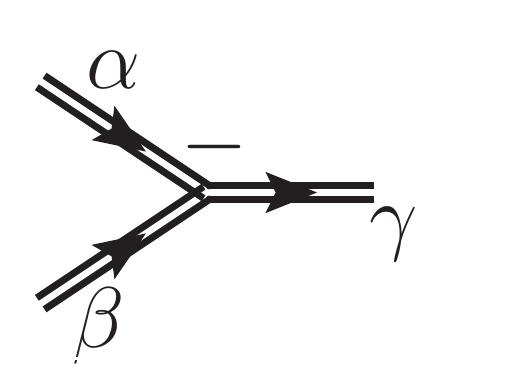}
}
\hspace{-3mm}
\equiv
\raisebox{-0.41\height}{
	\includegraphics[scale=0.45]{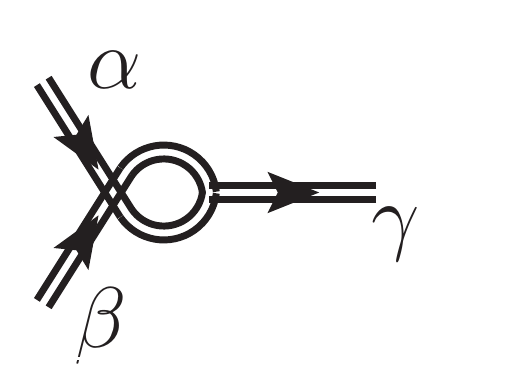}
}
\hspace{-3mm}
=
\sum_{a}{
\frac{
\hspace{-1.5mm}
\raisebox{-0.45\height}{
	\includegraphics[scale=0.3]{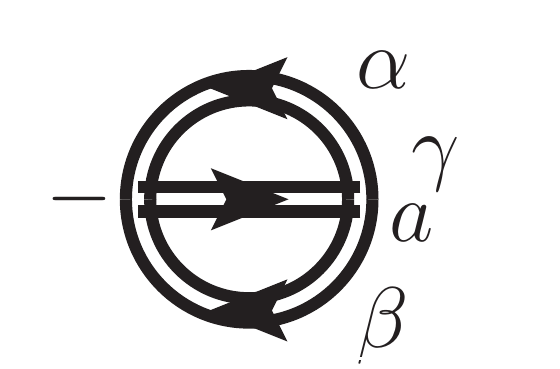}
}
\hspace{-3mm}
}{
\raisebox{-0.45\height}{
	\includegraphics[scale=0.3]{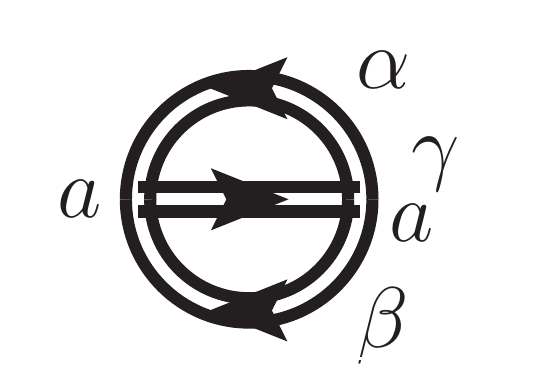}
}
\hspace{-3mm}
}
\raisebox{-0.43\height}{
	\includegraphics[scale=0.45]{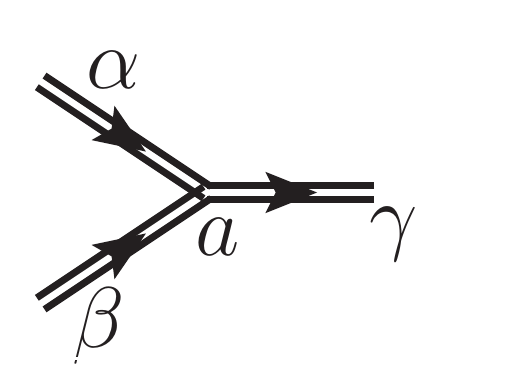}
}
}
\hspace{-3mm},
\end{equation}
where the equivalence defines Yutsis' notation \cite{YutsisNotation}.
Typically this just gives a sign $\pm1$, for example we have a minus sign
for the antisymmetric triple-gluon vertex.

\subsubsection{Example: $4\to 5$ gluons}

The method of evaluating the radiation matrices with the above stated
relations will first be applied to a $4 \to 5$ gluon example, and after
that a general formula will be derived.
Let us thus consider a four-gluon basis vector radiating a fifth
gluon from gluon 3. In diagrammatic form, 
denoting the standard triple gluon vertex, $i f^{abc}$, with
a black dot, where the indices are read in counter
clockwise order, the color structure becomes
\begin{equation}\label{eq:RadiationMatrixExample}
N^{\alpha_1}
\raisebox{-0.45\height}{
	\includegraphics[scale=0.5]{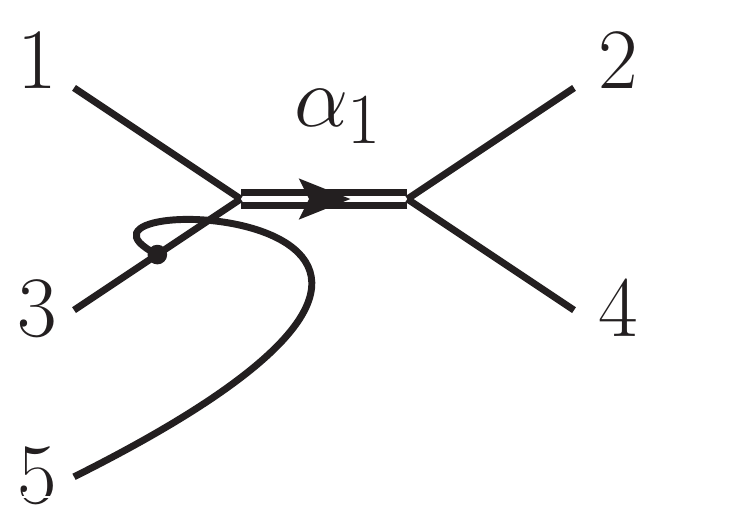}
}
\hspace{-1mm},
\end{equation}
where we have drawn the fifth gluon such that the gluon ordering
in \eqref{eq:MultipletBasisVector} is respected.

Applying the completeness relation \eqref{eq:CompletenessRelation}
to gluon 5 and the representation $\alpha_1$ gives
\begin{equation}\label{eq:RadiationMatrixExampleCR}
N^{\alpha_1}
\raisebox{-0.45\height}{
	\includegraphics[scale=0.45]{figures/RadiationMatrix/Example/InitialColorStructure}
}
=
-
\sum_{\beta_1\in{}\Adj \otimes\alpha_1}{
\frac{
	d_{\beta_1}
}{
	\includegraphics[scale=0.3]{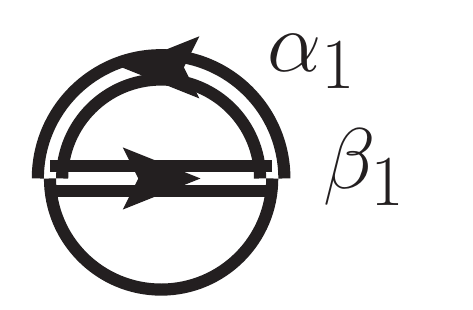}
	\hspace{-2mm}
}
N^{\alpha_1}
\raisebox{-0.45\height}{
	\includegraphics[scale=0.45]{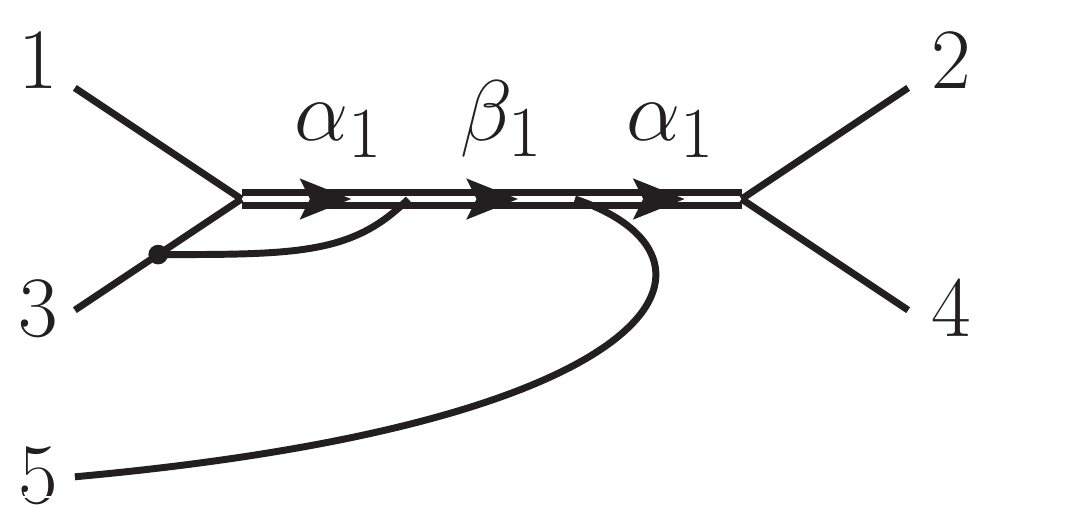}
}
}
\hspace{-1mm},
\end{equation}
where the sum runs over representations in the adjoint
representation times $\alpha_1$. On the right hand side above,
gluon 1 and 3 and the representation $\beta_1$ are connected
by a vertex correction.
Using \eqref{eq:ColorStructureVertexCorrection} with
$\gamma\rightarrow\beta_1$, $\epsilon\rightarrow\alpha_1$
and $\alpha,\beta,\delta,\zeta\rightarrow{}\Adj$ to remove it,
the color structure is
\begin{equation}\label{eq:RadiationMatrixExampleVertexCorrection}
N^{\alpha_1}
\raisebox{-0.45\height}{
	\includegraphics[scale=0.45]{figures/RadiationMatrix/Example/InitialColorStructure}
}
=
-
\sum_{ \substack{\beta_1\in{}(\Adj \otimes\alpha_1\cap\Adj\otimes{}\Adj)\\ a}}{
\frac{d_{\beta_1}}{
	\includegraphics[scale=0.3]{figures/RadiationMatrix/Example/Wig3jCR}
	\hspace{-2mm}	
}
\frac{
	\raisebox{-0.25\height}{
		\includegraphics[scale=0.45]{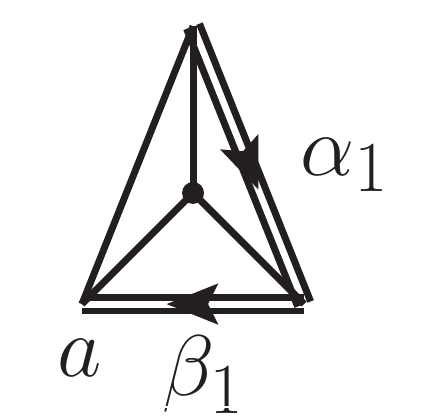}
	}
	\hspace{-3mm}
}{
        \hspace{-1.75mm}
	\includegraphics[scale=0.3]{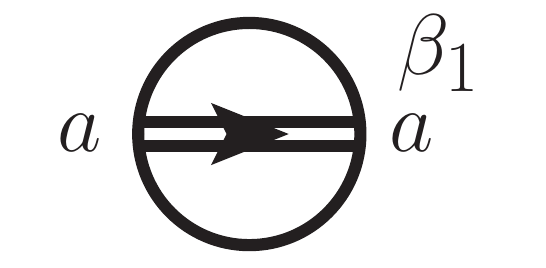}
	\hspace{-3mm}
}
\frac{N^{\alpha_1}}{N^{\beta_1\alpha_1}}
N^{\beta_1\alpha_1}
\raisebox{-0.45\height}{
	\includegraphics[scale=0.45]{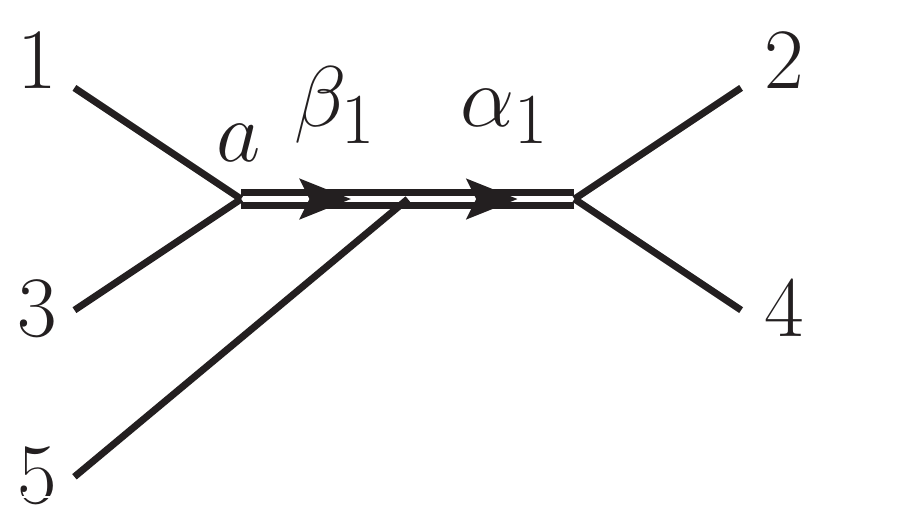}
}
}
\hspace{-2mm}.
\end{equation}
This equation is now of the desired form, \eqref{eq:RadiationMatrixDefinition}
where the radiation matrix components trivially
can be read off by comparison to \eqref{eq:MultipletBasisVector}.
Letting $\beta=(\beta_1,\beta_2)$ denote the representation set labeling the $5$-gluon basis vector,
we immediately see that the representation $\beta_2$ is constrained to be
$\beta_2=\alpha_1$.
Thus most of the projections onto the $5$-gluon basis vectors vanish.

Note that \eqref{eq:RadiationMatrixExampleVertexCorrection} has been derived
without explicitly stating the representation $\alpha_1$.
The result is thus generic and, knowing the Wigner coefficients
and the dimensions of the representations,
it can be used for immediately writing down the decomposition for any
initial $\Vec^{\alpha}$.
As an example, if $\alpha_1=10$ the allowed $\beta_1$
representations, i.e., those present in both $\Adj\otimes\alpha_1$ and $\Adj \otimes{}\Adj$,
are 8, 10, 27 and 0 (for $\Nc \geq 4$).
If $\beta_1=8$, there are two possible vertices connecting gluon 1 and 3
in the 5-gluon basis vector, and similarly if $\beta_1=10$
(and $\Nc\geq 4$) there are two vertices connecting $\beta_1$ and $\alpha_1$.

For evaluating the right hand side of
\eqref{eq:RadiationMatrixExampleVertexCorrection} we use the dimensions of
the representations, stated in \eqref{eq:RepresentationDimensions},
and the Wigner coefficients calculated as in \cite{Sjodahl:2015qoa}.
Ordering the allowed representations $\beta_1$ as
{$(8^{s}$, $8^{a}$, $27$, $10^{f}$, $10^{fd},0)$}, the Wigner $6j$ coefficient in \eqref{eq:RadiationMatrixExampleVertexCorrection} takes the values
\begin{equation}\label{eq:ExampleColumn6js}
\left(
\frac{-1}{\sqrt{ \Nc^2-4} \left(\Nc^2-1\right)},
0,
\frac{1}{\sqrt{ \Nc^2+3\Nc+2} \left(\Nc^2-\Nc\right)},
\frac{\sqrt{2}}{\sqrt{ \Nc^2-4} \left(\Nc^2-1\right)},
0,
\frac{1}{\sqrt{ \Nc^2-3\Nc+2} \left(\Nc^2+\Nc\right)}
\right)
\end{equation}
respectively.
In \cite{Sjodahl:2015qoa} the Wigner $3j$ coefficients are normalized to one.
Eq. (\ref{eq:RadiationMatrixExampleVertexCorrection}) is valid for
any normalization as long as it is consistently used. However,
requiring all $3j$ coefficient to be one implies a
normalization $i\tilde f^{abc}i\tilde f^{cba}=1$ for the triple-gluon
vertex.
To get the correct $if^{abc}$-normalization, \eqref{eq:ExampleColumn6js}
must be therefore multiplied by a factor of
\begin{equation}\label{eq:Wig3jStructureConstant}
\sqrt{
\hspace{-1mm}
\raisebox{-0.4\height}{
	\includegraphics[scale=0.3]{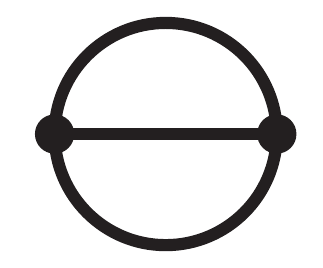}
}  }
=
\sqrt{2\Nc(\Nc^2-1)T_R}\;\; \mbox{  (using standard normalization of vertices)}.
\end{equation}
Normalizing the Wigner $3j$ coefficients to one and using the definition of the normalization constants, \eqref{eq:MultipletBasisVectorNormalizationFactor}, gives
\begin{equation}\label{eq:Wig3jNormalizationDifference}
\frac{N^{\alpha_1}}{N^{\beta_1\alpha_1}}
  =
  \frac{1}{\sqrt{d_{\beta_1}}}.
\end{equation}
Combining the dimensions, the Wigner $6j$ coefficients, the normalization factor
\eqref{eq:Wig3jStructureConstant}, and the overall sign in
\eqref{eq:RadiationMatrixExampleVertexCorrection} gives
\begin{equation}\label{eq:ExampleColumnTotal}
\sqrt{\TR}
\left(
\sqrt{\frac{2\Nc}{ \Nc^2-4}},
0,
-\sqrt{\frac{\Nc(\Nc+3)}{2(\Nc+2)}},
-\sqrt{\Nc},
0,
-\sqrt{\frac{\Nc(\Nc-3)}{2(\Nc-2)}}
\right).
\end{equation}
These are the factors required to express the color structure of
\eqref{eq:RadiationMatrixExample} in terms of the five-gluon basis
vectors.
For bases which are not charge conjugation invariant, they would be the
entries in the radiation matrix $\mathbf{T}_{3}$, corresponding to mapping the
initial vector $\Vec^{10}$ emitting a gluon $g_5$ from gluon $g_3$
onto the five-gluon basis vectors
$\Vec^{8s,10}$,
$\Vec^{8a,10}$,
$\Vec^{27,10}$,
$\Vec^{10,10f}$,
$\Vec^{10,10fd}$ and
$\Vec^{10,0}$.
In the above case, the initial color structure,
\eqref{eq:RadiationMatrixExample} with $\alpha_1=10$,
is from a basis vector which is not charge conjugation invariant,
and neither is any of the vectors which the color structure is
projected onto.
To get to the charge conjugation invariant vectors in this case
simply requires the substitution of $10 \to 20$ in the vector
representation labels. With this change, the result, \eqref{eq:ExampleColumnTotal},
can be compared to column five of $\mathbf{T}_{3}$
in \eqref{eq:RadiationMatrix_4g_to_5g_Gluon_3} in \appref{sec:5g},
where the radiation matrices for $4\to5$ gluons are given
expressed in the basis from \eqref{eq:MultipletBasisVectors5g}
(which is also electronically attached as an online resource).

In general, when knowing the radiation matrices in non-charge conjugation
invariant bases and converting to charge conjugation invariant bases,
a sign might be required since the $\Ng$-gluon
non-invariant vector may come with a minus sign in the linear combination
building up the charge conjugation invariant vector.
For the same reason another minus sign may come from the $(\Ng-1)$-gluon
basis vector.
Apart from a potential sign, factors compensating for vector normalizations
and occurrence of both a vector and its conjugate on the right hand side
of \eqref{eq:RadiationMatrixDefinition} may be required.

For five external gluons (and $\Nc \geq 4$) 
there are 22 charge conjugation invariant basis vectors, stated in
\eqref{eq:MultipletBasisVectors5g}, but the example color structure
is given by only four of them. It is worth noting that although, in
this case, most of the basis vectors with $\beta_2=10$ contribute
(there are only two zeros in \eqref{eq:ExampleColumnTotal}),
for more external gluons the constraints corresponding to
$\beta_1\in{}\alpha_1\otimes{}\Adj$ become more restrictive.
This point will be elaborated on after the derivation of the general
formula for radiation matrices.

\subsubsection{The general case: $n-1\to n$ gluons}

In general, if the $n$th gluon is radiated from one of the incoming gluons,
the color structure of \eqref{eq:radiation_color} will be of the form
\begin{equation}\label{eq:RadiationMatrixDerivationInitialVector}
N^{ \alpha_1 \, \alpha_2\dots\alpha_{\Ng-1-3}}
\raisebox{-0.45\height}{
	\includegraphics[scale=0.6]{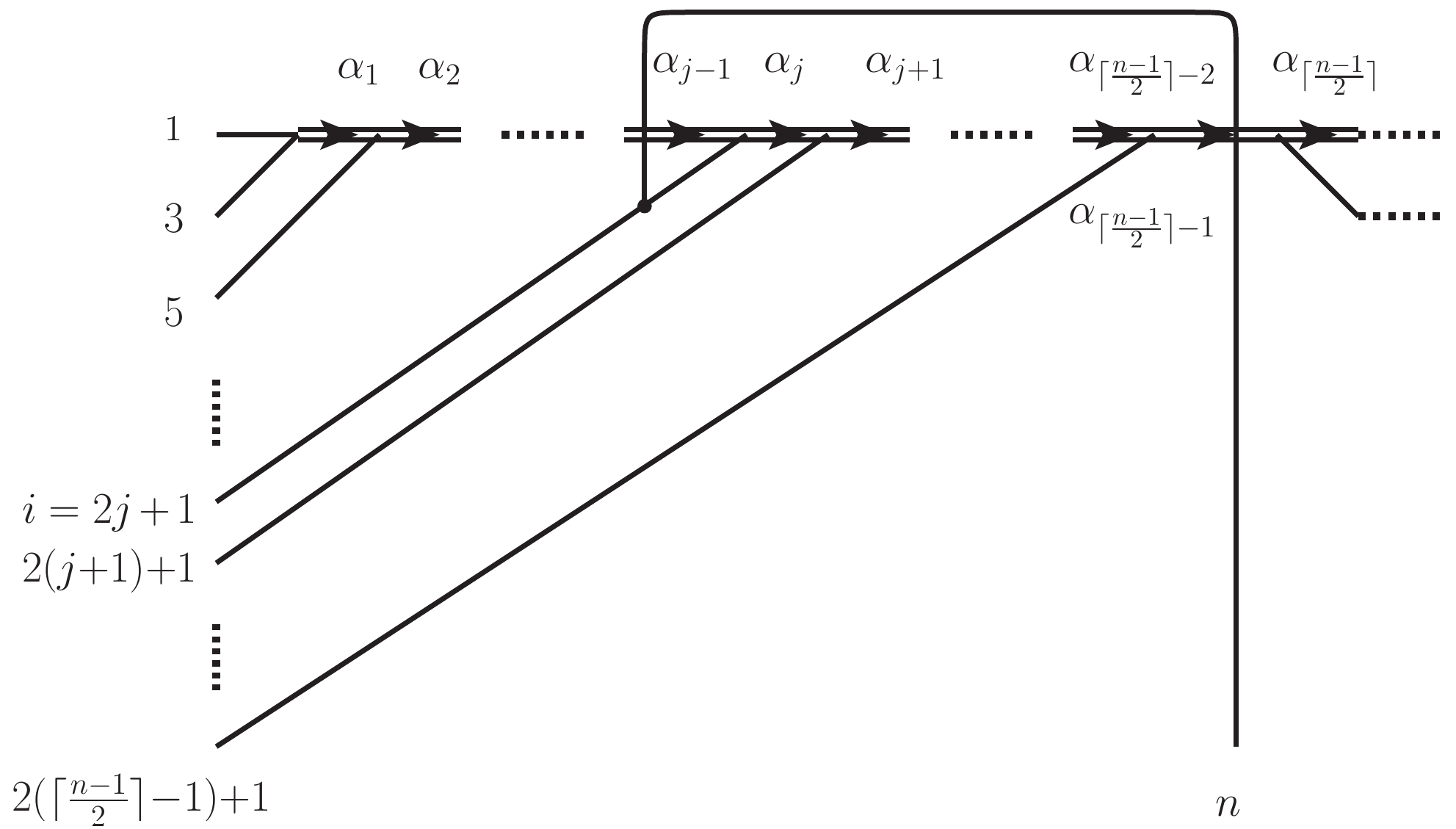}
},
\end{equation}
where, if $j=0$ or 1 (i.e. $i=1$ or 3), $\alpha_{j-1}$ is an octet. In the following steps it will be assumed that gluon 1 is not the emitter, this special case will be addressed after the derivation. To get to the right hand side of
\eqref{eq:RadiationMatrixDefinition}, the same steps as in the above example are used:
First the completeness relation, \eqref{eq:CompletenessRelation},
is applied repeatedly and then vertex corrections are removed
using \eqref{eq:ColorStructureVertexCorrection}.

We thus want to apply the completeness relations to gluon $\Ng$ and the representations
$\alpha_{j},\alpha_{j+1},\dots{},\alpha_{\lceil\frac{n-1}{2}\rceil-1}$, i.e.,
we insert it in the encircled positions in
\begin{equation}\label{eq:RadiationMatrixDerivationCRPlacements}
  N^{\alpha_1 \, \alpha_2\dots\alpha_{\Ng-4}}
  \raisebox{-0.45\height}{
    \includegraphics[scale=0.6]{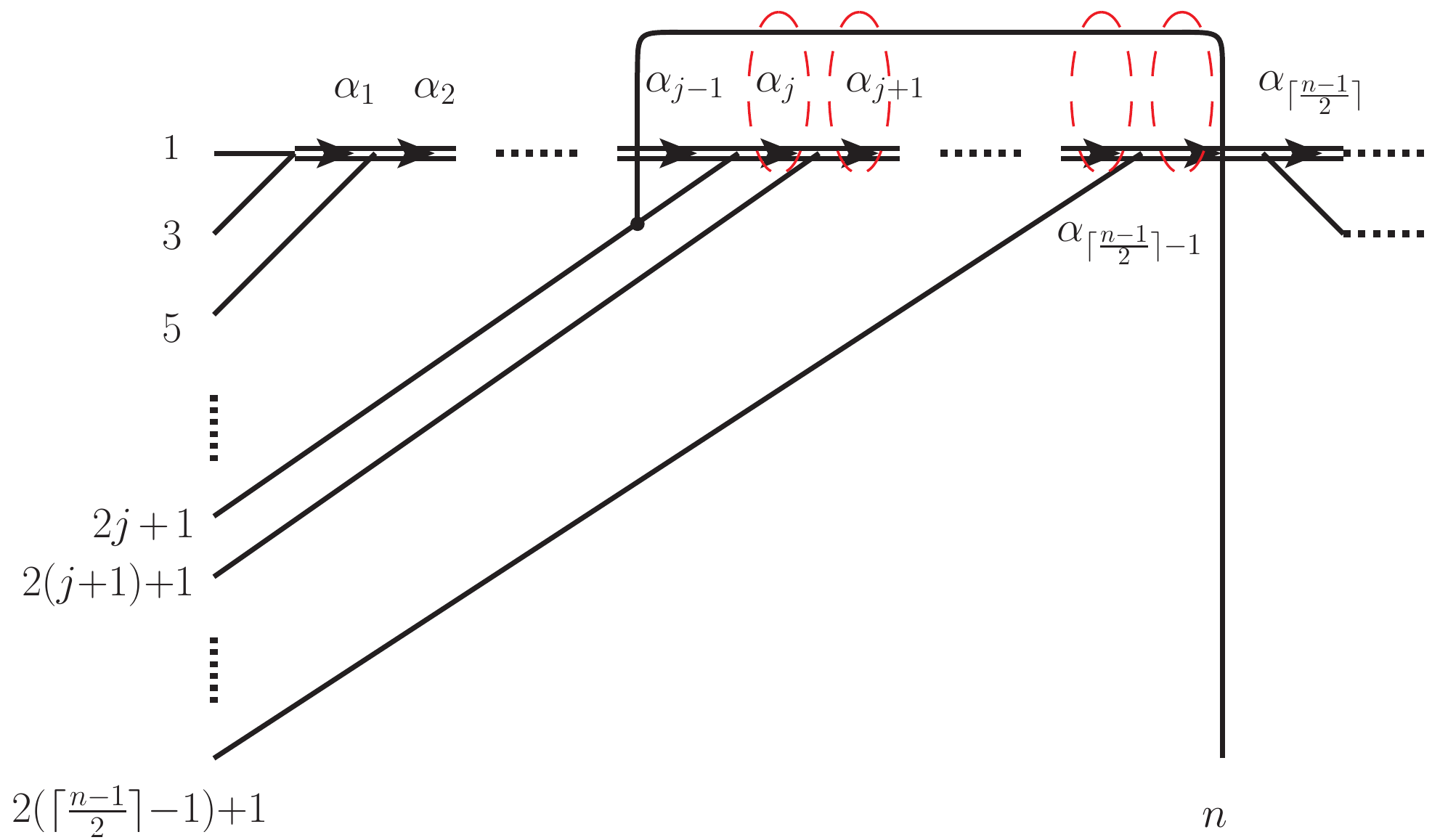}
  },
\end{equation}
resulting in\newpage
$$
\sum_{
	\beta_{j},\beta_{j+1},\dots,
	\beta_{\lceil\frac{n-1}{2}\rceil-1}
}
{
\frac{d_{\beta_{j}}}
     {\raisebox{-0.45\height}{
	 \includegraphics[scale=0.3]{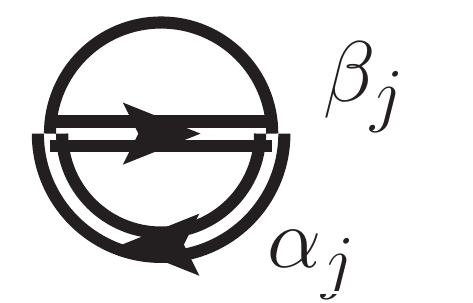}}}
\frac{d_{\beta_{j+1}}}
     {\raisebox{-0.45\height}{
	 \includegraphics[scale=0.3]{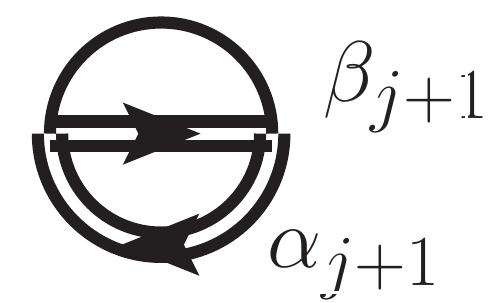}}}
\dots
\frac{d_{\beta_{\lceil\frac{n-1}{2}\rceil-1}}}
     {\raisebox{-0.45\height}{
	 \includegraphics[scale=0.3]{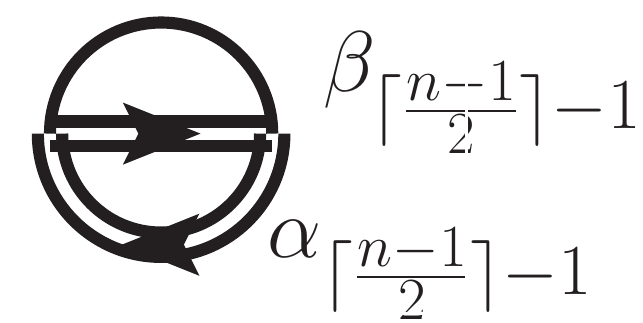}}}
 N^{\alpha_1 \, \alpha_2\dots\alpha_{\Ng-4}}
}
$$
\begin{equation}\label{eq:RadiationMatrixDerivationCRApplied}
\times
\raisebox{-0.45\height}{
	\includegraphics[scale=0.6]{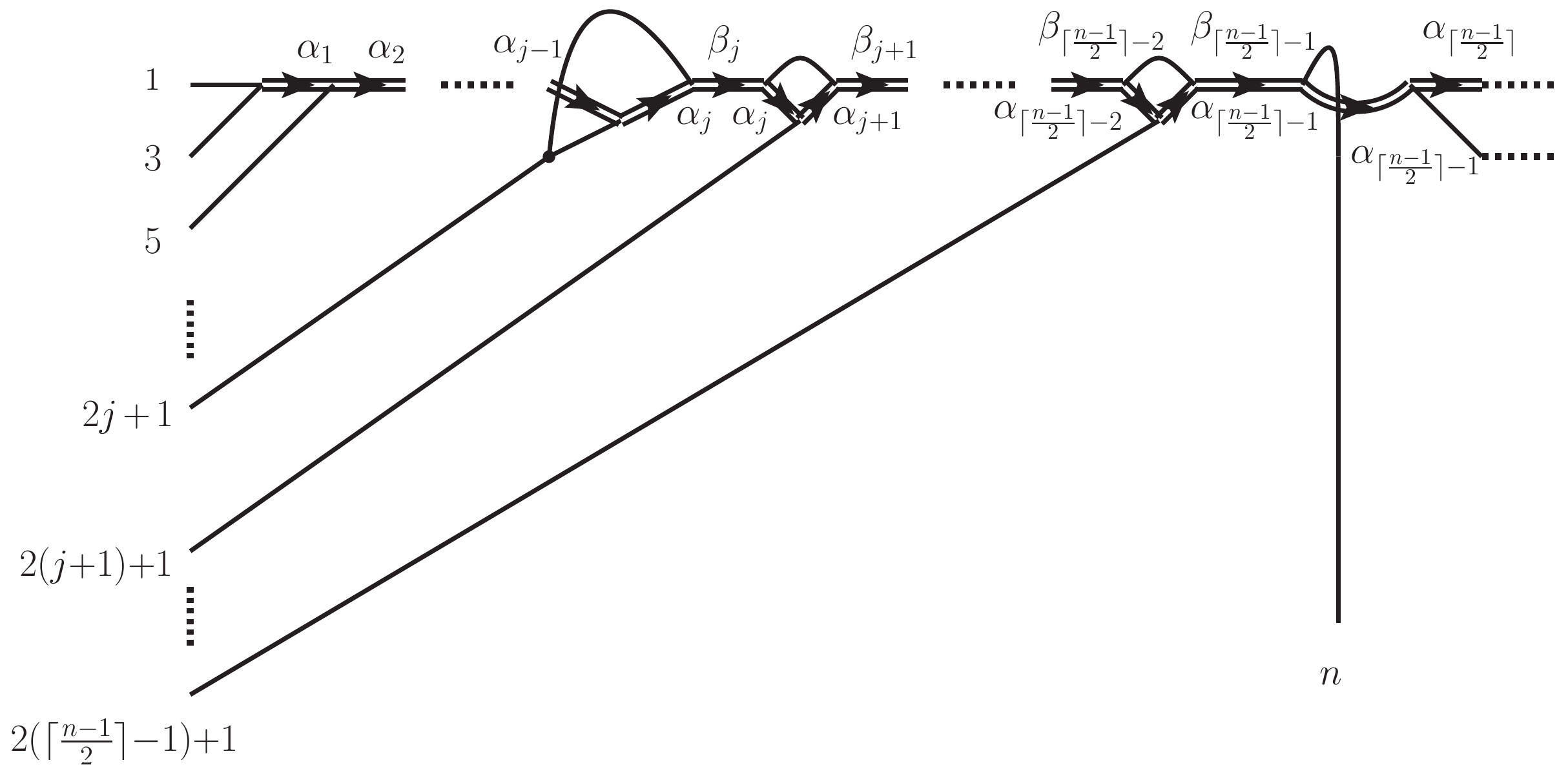}
}.
\end{equation}
Here gluon $2j+1$ has a vertex correction of a different form
compared to the other gluons. We first treat this separately
and then address all other vertex corrections.
Using Yutsis' notation, \eqref{eq:YutsisVertexToNormalVertex},
and \eqref{eq:ColorStructureVertexCorrection},
the leftmost vertex correction can be written
\begin{equation}\label{eq:RadiationMatrixDerivationVertexCorrectionSpecial}
\raisebox{-0.45\height}{
	\includegraphics[scale=0.6]{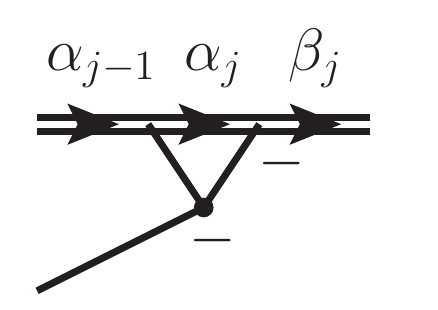}
}
=
\sum_{ b_j
}
{
\hspace{3mm}
\frac{
\hspace{-3.5mm}
\raisebox{-0.45\height}{
	\includegraphics[scale=0.45]{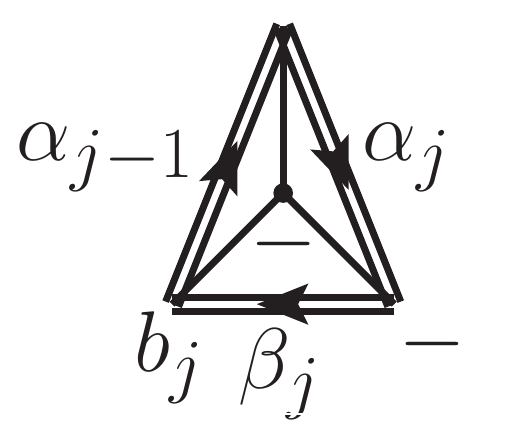}
}
\hspace{-3mm}
}{
\raisebox{-0.45\height}{
	\includegraphics[scale=0.3]{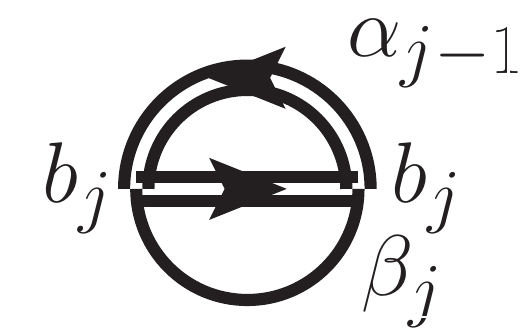}
}
\hspace{-2mm}
}
\raisebox{-0.45\height}{
	\includegraphics[scale=0.6]{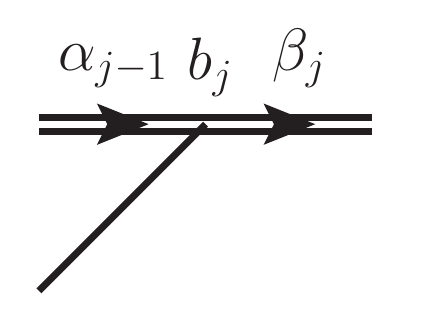}
}
}
.
\end{equation}
The remaining vertex corrections can be contracted similarly, for example
\begin{equation}\label{eq:RadiationMatrixDerivationVertexCorrection}
\raisebox{-0.45\height}{
	\includegraphics[scale=0.6]{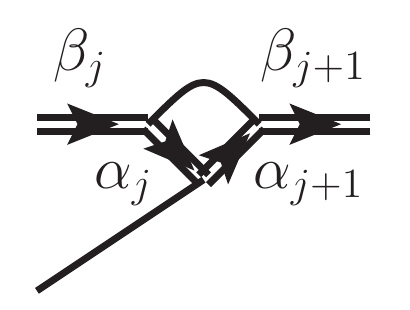}
}
=
\sum_{ b_{j+1}
}
{
\frac{
\hspace{1mm}
\raisebox{-0.45\height}{
	\includegraphics[scale=0.45]{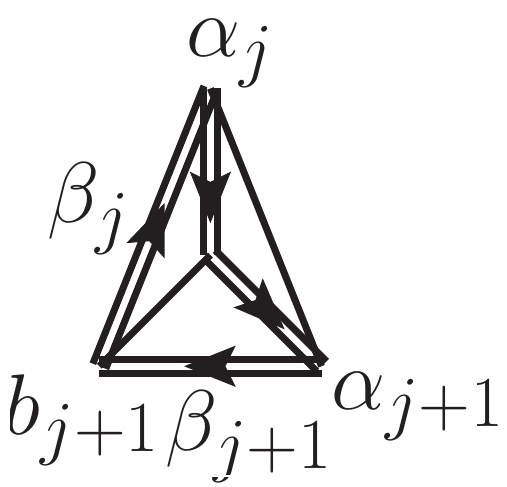}
}
\hspace{-3mm}
}{
\raisebox{-0.45\height}{
	\includegraphics[scale=0.3]{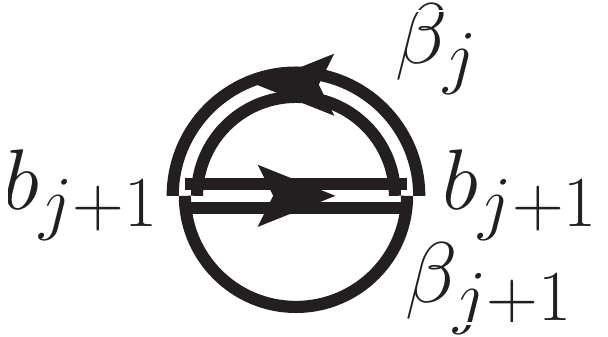}
}
\hspace{-1mm}
}
\raisebox{-0.45\height}{
	\includegraphics[scale=0.6]{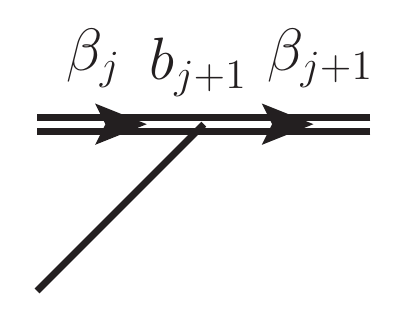}
}
}
.
\end{equation}
Applying the above steps and adjusting the vertex order of the last new
vertex results in
$$
\sum_{
  \substack{
	\beta_{j},\beta_{j+1},\dots,
	\beta_{\lceil\frac{n-1}{2}\rceil-1}\\
        b_{j},b_{j+1},\dots,
	b_{\lceil\frac{n-1}{2}\rceil}
  }
}{
\frac{d_{\beta_{j}}}{
	\raisebox{-0.45\height}{
		\includegraphics[scale=0.3]{figures/RadiationMatrix/GeneralCase/Wig3j/Wig3j_j}
	}
	\hspace{-4mm}}
\frac{
	\hspace{-6mm}
	\raisebox{-0.25\height}{
		\includegraphics[scale=0.4]{figures/RadiationMatrix/GeneralCase/VertexCorrection/Wig6j_Leg2jP1}
	}
	\hspace{-5mm}
}{
	\hspace{-1mm}
	\raisebox{-0.30\height}{
		\includegraphics[scale=0.3]{figures/RadiationMatrix/GeneralCase/VertexCorrection/Wig3j_Leg2jP1}
	}
	\hspace{-2mm}
}
\left[
\prod_{k=j+1}^{\lceil\frac{n-1}{2}\rceil{-1}}{
\frac{d_{\beta_{k}}}{
	\raisebox{-0.45\height}{
		\includegraphics[scale=0.3]{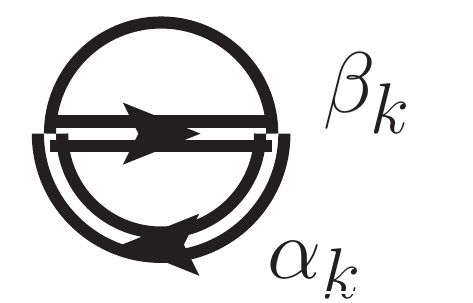}
	}
	\hspace{-3mm}
}
\frac{
	\hspace{-6mm}
	\raisebox{-0.25\height}{
		\includegraphics[scale=0.4]{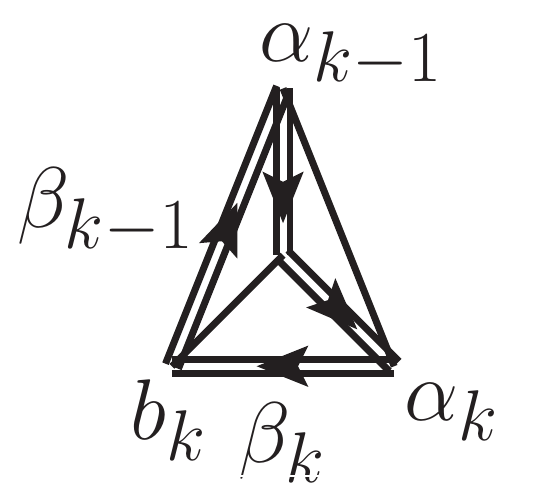}
	}
	\hspace{-2mm}
}{
	\hspace{-2mm}
	\raisebox{-0.30\height}{
		\includegraphics[scale=0.3]{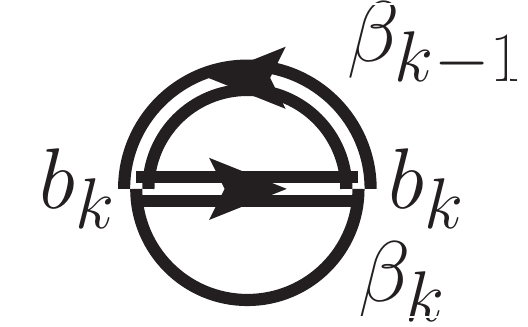}	
	}
	\hspace{-3mm}
}
}
\right]
\frac{
	\hspace{-2mm}
	\raisebox{-0.1\height}{
		\includegraphics[scale=0.3]{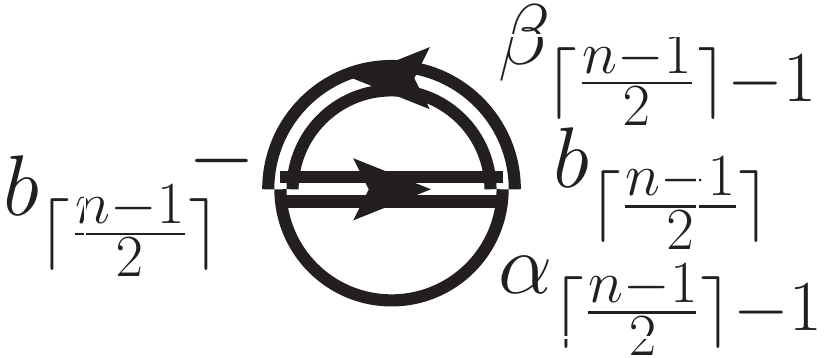}
	}
	\hspace{-3mm}
}{
	\hspace{1mm}
	\raisebox{-0.30\height}{
		\includegraphics[scale=0.3]{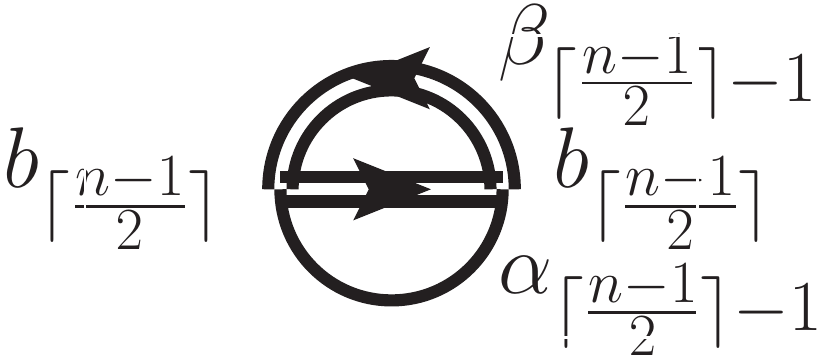}
	}
	\hspace{-3mm}
}
}
$$
$$
\times
\frac{
	N^{\alpha_1 \, \alpha_2\dots\alpha_{\Ng-4}}
}{
	N^{\alpha_1 \dots \alpha_{j-1}\beta_{j}\dots\beta_{\lceil\frac{n-1}{2}\rceil{-1}}\,\alpha_{\lceil\frac{n-1}{2}\rceil{-1}}\dots\alpha_{\Ng-4}}
}
N^{\alpha_1 \dots \alpha_{j-1}\beta_{j}\dots\beta_{\lceil\frac{n-1}{2}\rceil{-1}}\,\alpha_{\lceil\frac{n-1}{2}\rceil{-1}}\dots\alpha_{\Ng-4}}
$$
\begin{equation}\label{eq:RadiationMatrixDerivationFinal}
\times
\raisebox{-0.45\height}{
	\includegraphics[scale=0.6]{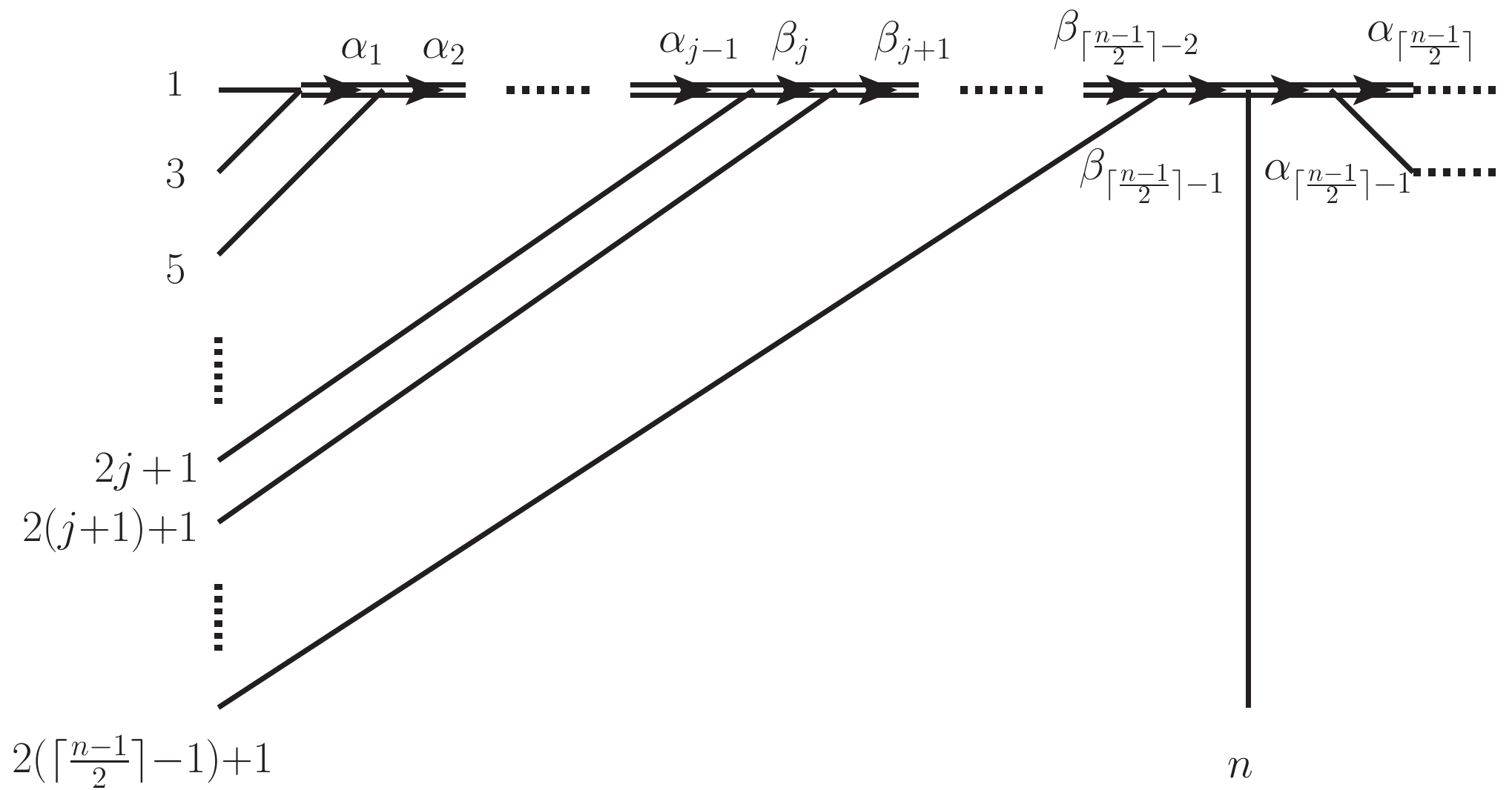}
},
\end{equation}
where the vertex labels in the third line have been suppressed.
The third line of \eqref{eq:RadiationMatrixDerivationFinal}, combined with the last normalization factor of the second line is now of the form of a basis vector. Hence the factor in front of it is the radiation matrix $\mathbf{T}_{2j+1}$ expressed in terms of Wigner coefficients.

We remark that the form of the radiation matrix from the example,
\eqref{eq:RadiationMatrixExampleVertexCorrection}, differs from
\eqref{eq:RadiationMatrixDerivationFinal}. This is only due to the fifth gluon being drawn in an, at that point, more natural way. The two expressions are identical, if the expression from \eqref{eq:RadiationMatrixDerivationFinal} is written out for gluon 3 with $n=5$, it can be simplified to become exactly the expression of \eqref{eq:RadiationMatrixExampleVertexCorrection}.

The derived result, \eqref{eq:RadiationMatrixDerivationFinal}, is for gluons emitted from the ``incoming'' gluons. For gluons emitted from the ``outgoing'' gluons an analogous derivation can be done, resulting in an equation similar to \eqref{eq:RadiationMatrixDerivationFinal}.
A special case occurs if the emitter is the first gluon on its side,
gluon 1 for the ``incoming" side and gluon 2 for the ``outgoing" side.
Compared to the radiation matrices $\mathbf{T}_{3}$ and $\mathbf{T}_{4}$
the matrices $\mathbf{T}_{1}$ and $\mathbf{T}_{2}$ are identical up to
sign differences for some entries. The difference originates from a difference in the leftmost (rightmost for gluon 2) vertex correction,  \eqref{eq:RadiationMatrixDerivationVertexCorrectionSpecial}, that results in a change in vertex ordering which gives a possible sign difference.
(Taking gluon $1$ and gluon $\Ng$ to be the gluons with shifted momenta,
emission from gluon 1 need not be considered, see \figref{fig:divisions}.)

Concerning the group theoretical constraints on the representations we note that each representation to the left of $\beta_j$ and to the right of $\beta_{\lceil\frac{n-1}{2}\rceil-1}$ are constrained to be equivalent to representations in the set $\alpha$, i.e.,
\bea
\beta_{k}&=&\alpha_{k},\;\;\;\;\mathrm{for}\;k=1,\dots,j-1,\nn
\beta_{k}&=&\alpha_{k-1},\;\mathrm{for}\;k=\lceil\frac{n-1}{2}\rceil,\dots,n-3.
\label{eq:UnchangedRepresentations}
\eea
The shift in the index of the second equality in \eqref{eq:UnchangedRepresentations} is from the fact that there are $n-3$ representations for $n$ gluons and $n-4$ for $n-1$ gluons, with the new representation being inserted just before $\alpha_{\lceil\frac{n-1}{2}\rceil-1}$.

There are also constraints on the set of representations $\beta$ coming from the completeness relations applied in \eqref{eq:RadiationMatrixDerivationCRApplied},
\begin{equation}\label{eq:RepresentationConstraints}
\beta_{k}\in{}\alpha_{k}\otimes{}\Adj,\;\mathrm{for}\;k=j,j+1,\dots,\lceil\frac{n-1}{2}\rceil-1.
\end{equation}
This constraint will become more restrictive when the emitting gluon
is far from the middle of the basis vector, since in this case,
the constraint is imposed on many representations in the basis vector.

Using the constraints \eqref{eq:UnchangedRepresentations}
and \eqref{eq:RepresentationConstraints}, the number of possibly
non-zero elements in the radiation matrix columns can be counted
(more are zero due to generalized vertices giving Wigner
coefficients which are zero). The result, averaged over
all possible emitters $2,..,\Ng-1$ and all possible initial basis vectors,
is shown in \tabref{tab:RadiationMatrixNonZero} along with the
maximal number of possible $\beta$ for any $\alpha$.
Both the maximal number of possible $\beta$ for any $\alpha$ and the
average over all $\alpha$ and all emitters are overestimates, as
symmetries of the Wigner coefficients will force some of them to vanish,
depending on the choice of vertices.
In addition, there are Wigner coefficients vanishing due to the invariance condition of tensors under the group, see \cite{Cvi08} for the invariance condition in birdtrack notation.
For the calculated radiation matrices, the reductions due to vanishing Wigner coefficients changes the averages for $\Nc\geq{}n$ from $6.9$ to $3.9$ and $14.4$ to $8.9$ for the $n=5$ and $n=6$ cases, respectively. For the gluons only case there is the further reduction due to charge conjugation invariance.
For comparison, \tabref{tab:RadiationMatrixNonZero}
also shows the dimensions of the (all order) vector space,
both for QCD and in the limit $N_c\rightarrow\infty$,
and the number of vectors in the spanning sets for the ``trace bases'' and ``DDM bases''.

\begin{table}
\centering
\begin{tabular}{ |c|c|r|r|r|r|r|r|r|r|r|}
\hline
 				& $\Ng$	             &5 &$6$           & $7$ 	& $8$     & $9$         & $10$     & $11$ &    $12$\\
\hline
Avg				& QCD 		     &  6.0     & 10.8  &  17.5   &  32.5       &   54.6   & 106       & 185        & 268\\
				& $\Nc\geq{}\Ng$     &    6.9   & 14.4  &  24.6   &  57.9       &    109   & 299       &593         & 1 775 \\
\hline
Max				& QCD               &    8     & 33    &   33    & 178 	& 178      & 962       & 962        & 5 220\\

                                &$\Nc\geq{}\Ng$     &    9     & 44    &   44    & 400 	& 400      & 4 006     & 4 006       & 41 256 \\
\hline
Vectors                         & QCD	            &   32     & 145	&   702   & 3 598       & 19 280   &  107 160  & 614 000    & 3 609 760\\
(all orders) 			& $\Nc\geq{}\Ng$    &   44     & 265	& 1 854   & 14 833	& 133 496  & 1 334 961 & 14 684 570 &  176 214 841
\\
\hline
Trace  & any $\Nc$			             &   24     & 120   & 720     &       5 040 &   40 320 &   362 880 & 3 628 800 & 39 916 800
\\
\hline
DDM  & any $\Nc$			             &    6     &  24   & 120     &     720     &     5 040 &    40 320 &  362 880 & 3 628 800 \\
\hline
\end{tabular}
\caption{
The average (taken over all initial vectors and the emitters $2,...,\Ng-1$)
and maximal number of non-zero elements in the columns of the
radiation matrices for $(\Ng -1)\to \Ng$ gluons.
The stated numbers are overestimates since they assume that no
Wigner $6j$ coefficient involving admissible representations vanishes.
For comparison we also show
the total dimension of the all order vector space for $\Nc \ge \Ng$
and for $\Nc=3$ (without accounting for reduction due to charge
conjugation invariance) and the number of spanning vectors in the
tree-level trace bases and the DDM bases.}
\label{tab:RadiationMatrixNonZero}
\end{table}

We note that although the average (and maximal) number of terms
does increase with the number of gluons, compared to the increase
in the dimension of the vector space, this growth is very mild,
meaning that a smaller and smaller fraction of all basis vectors
contribute.
Instead of having to treat the square of the number of basis vectors
in the squaring of the amplitude, we thus only need to treat
the number of basis vectors, times the number of contributing
emitters times the average number of terms in deriving the amplitude;
recall that the squaring of the color structure itself is quick in the orthogonal
bases.
For example, using the trace basis for 10 gluons requires
$(362\, 880)^2 \sim 10^{11}$ terms in the squaring step, with the
(gluon specific) DDM basis this can be reduced to $\sim 10^9$,
whereas using the multiplet basis would require up
to $19\, 208\times 8\times 106\sim 10^7$ terms for
the identification of new basis vectors.
For more gluons the difference is
even larger. As our long term goal is to include processes
with an arbitrary number of quarks, we view the comparison
between the non charge conjugation invariant multiplet basis
and the trace basis as the most relevant comparison.
From \tabref{tab:RadiationMatrixNonZero} we thus conclude that the
overall treatment of the color structure can be sped up significantly
by the usage of multiplet bases.

\subsection{The five-gluon amplitude}
\label{sec:five-point}

\begin{figure}
  \centering
  \includegraphics[scale=0.5]{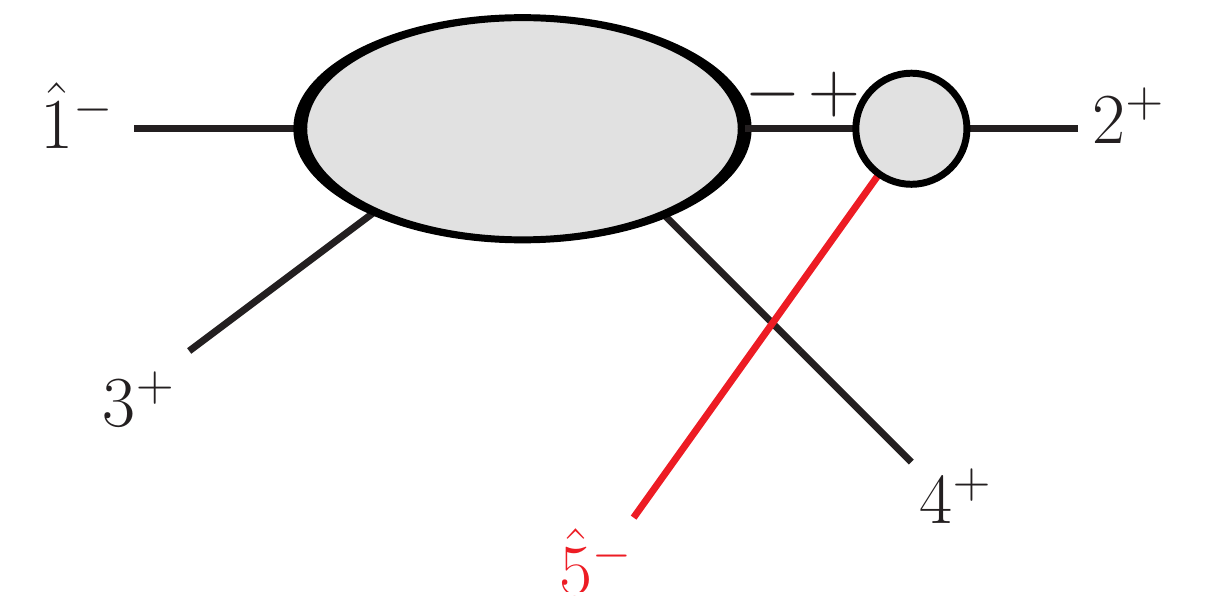}\quad
  \includegraphics[scale=0.5]{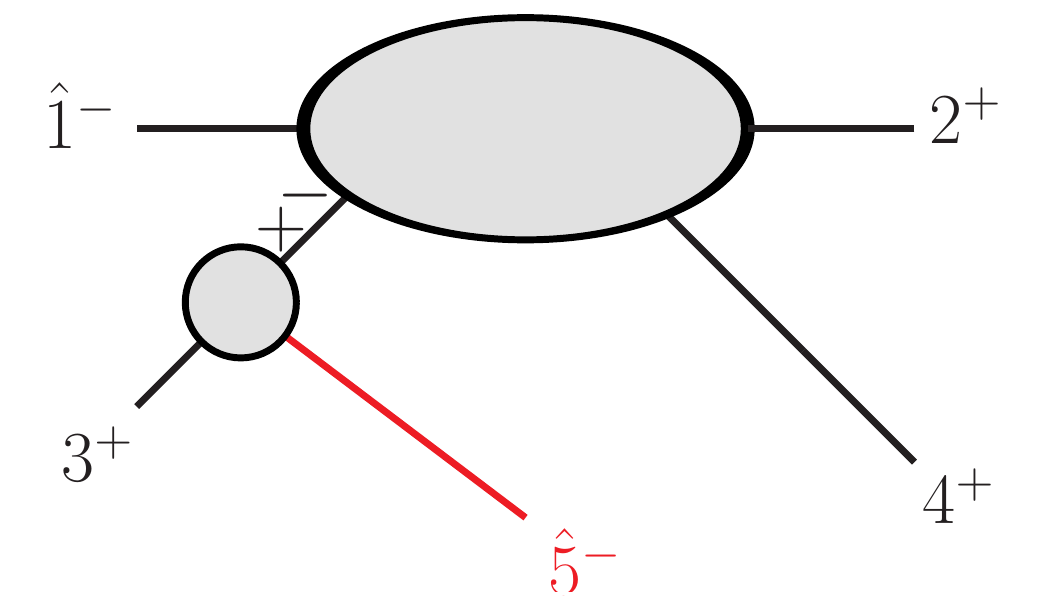}\quad	
  \includegraphics[scale=0.5]{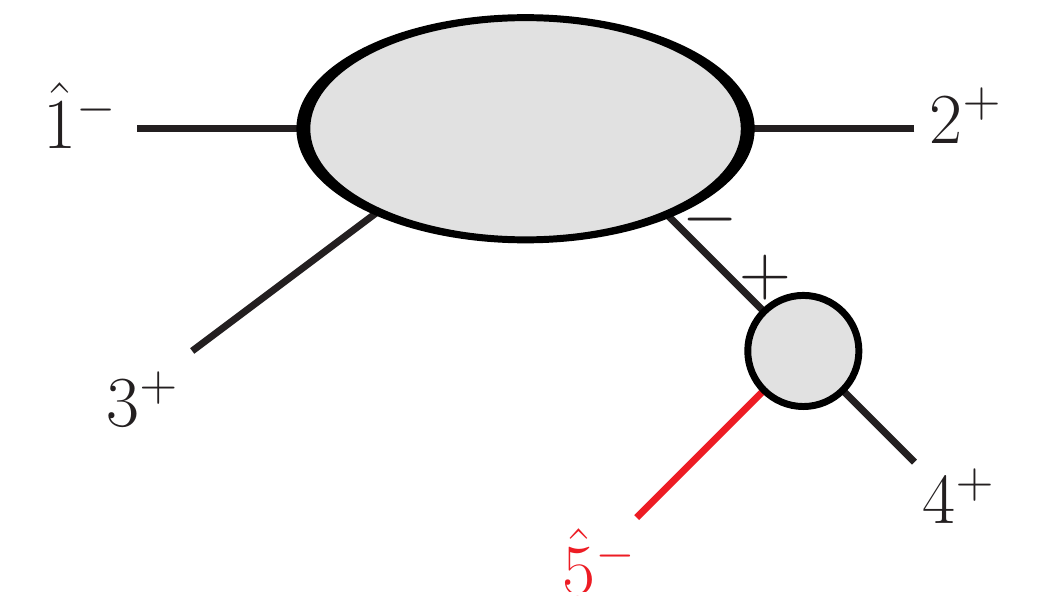}
\\
  \includegraphics[scale=0.5]{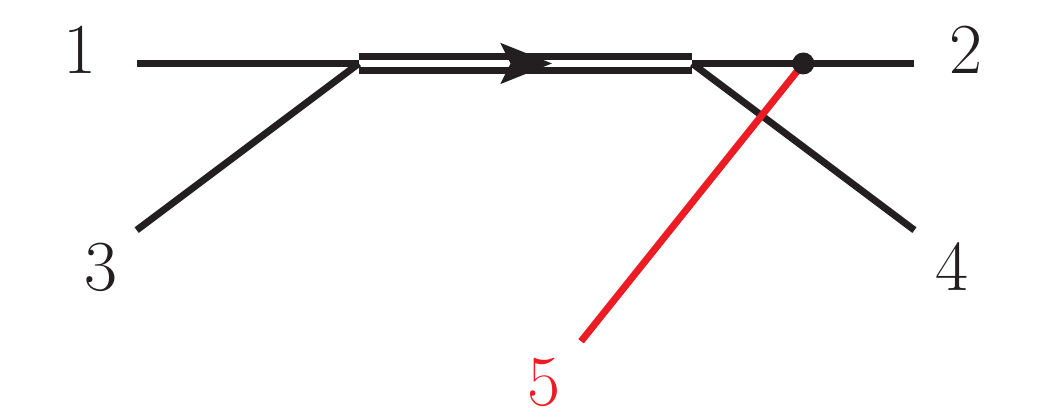}\quad \quad
  \includegraphics[scale=0.5]{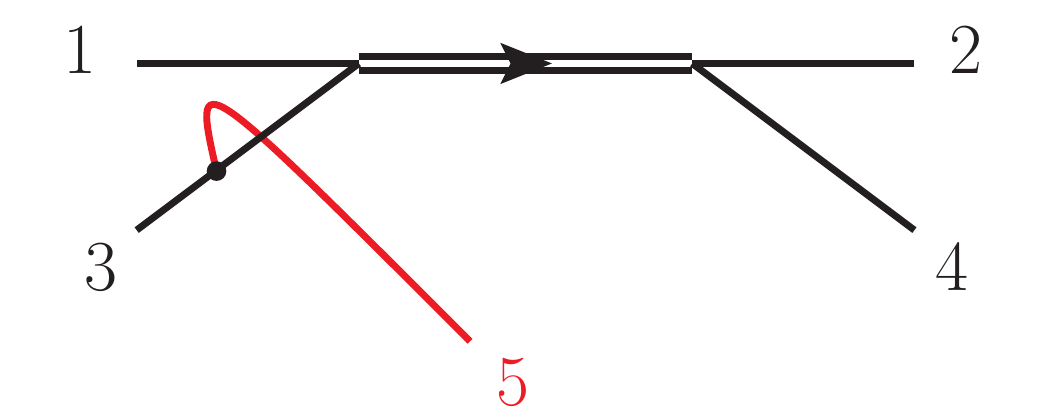}\quad \quad
  \includegraphics[scale=0.5]{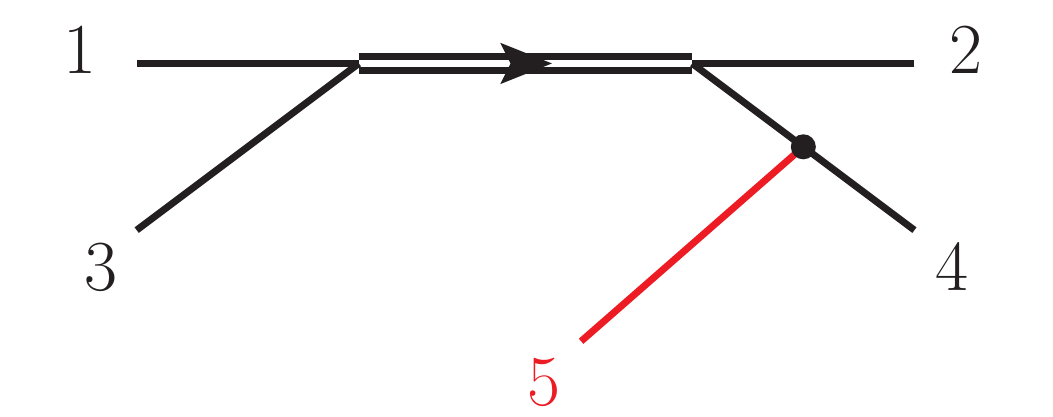}
  \caption{The relevant splittings for the BCFW recursion in the five-gluon MHV case (above) and
  the corresponding color structure for the radiation matrices (below). Note that the orders of the $if^{g_{i}g_{i,n}g_{n}}$ vertices are drawn to be consistent with \eqref{kinematic-recursion-MHV}.
} \label{fig:5pt-radiation}
\end{figure}

Utilizing that only the $(n-1,3)$ division contributes,
the BCFW recursion expression, \eqref{Color-dressed-BCFW-MHV}, for the
five-gluon color-dressed MHV amplitude is given by the sum of the
diagrams in \figref{fig:5pt-radiation}.
As seen in \eqref{4pt-kinematic-part}, the four-gluon color-dressed
(MHV) sub-amplitudes can be expressed in the multiplet basis.

The contractions of basis vectors in the four-gluon sub-amplitudes
and the structure constants in the three-gluon sub-amplitudes can
be expanded in the five-gluon multiplet basis using the radiation
matrices $\mathbf{T}_{i}$ for emitting the gluon $g_5$, illustrated in
\figref{fig:5pt-radiation}.
Collecting the kinematic factors corresponding to a given five-gluon
basis vector, we can write down the recursion relation
\eqref{kinematic-recursion-MHV} for the kinematic factor for five gluons
\bea
    {A}^{\beta}({1^-,2^+,3^+,4^+,5^-})&=&
    ~~\Sl_{\alpha}
  \left(\mathbf{T}_{2}\right)_{\beta \alpha}
   {A}^{\alpha}\left(\WH 1^-,\WH P^-_{2,5},3^+,4^+\right)
   {1\over s_{25}}
   {1\over\sqrt{\TR}}{\Spbb{2 \mid -\WH P_{2,5}}^3\over\Spbb{2\mid \WH 5}\Spbb{\WH 5\mid -\WH P_{2,5}} }\nn
   &&+\Sl_{\alpha}
    \left(\mathbf{T}_{3}\right)_{\beta \alpha}
   {A}^{\alpha}\left({\WH 1^-,2^+,\WH P^-_{3,5},4^+}\right)
   {1\over s_{35}}
   {1\over\sqrt{\TR}}{\Spbb{3 \mid -\WH P_{3,5}}^3\over\Spbb{3\mid \WH 5}\Spbb{\WH 5\mid -\WH P_{3,5}} }\nn
   &&+\Sl_{\alpha}
    \left(\mathbf{T}_{4}\right)_{\beta \alpha}
         {A}^{\alpha}\left(\WH 1^-,2^+,3^+,\WH P^-_{4,5}\right)
         {1\over s_{45}}
         {1\over\sqrt{\TR}}{\Spbb{4 \mid -\WH P_{4,5}}^3\over\Spbb{4\mid \WH 5}\Spbb{\WH 5\mid -\WH P_{4,5}} }
         .~~\Label{5pt-kinematic-1}
\eea
Here the matrices $\mathbf{T}_{i}$ in the first, second and third term
of \eqref{5pt-kinematic-1} are the radiation matrices corresponding to the
first, second and the third diagram in \figref{fig:5pt-radiation}.
They are calculated as in \secref{sec:color_recursion} and explicit
matrices are stated in appendix \ref{sec:5g}.
The four-gluon kinematic factors
${A}^{\alpha}\left(\WH 1^-,\WH P^-_{2,5},3^+,4^+\right)$,
${A}^{\alpha}\left({\WH 1^-,2^+,\WH P^-_{3,5},4^+}\right)$
and
${A}^{\alpha}\left(\WH 1^-,2^+,3^+,\WH P^-_{4,5}\right)$
are given by replacing the gluons $2$, $3$ and $4$ in \eqref{4pt-kinematic-part}
by $\WH P^-_{2,5}$, $\WH P^-_{3,5}$ and $\WH P^-_{4,5}$ respectively,
and inserting the explicit expressions for the color-ordered MHV
amplitudes \eqref{MHV-amplitude}. Using  the Mathematica package S@M  \cite{Maitre:2007jq}, one can compute the kinematic factors.
Clearly, in the five-gluon case, the only non-vanishing helicity
configurations are the MHV and the $\overline{\text{MHV}}$
making our result applicable to all relevant cases.

As shown in \secref{sec:comparison}, one can always express the
kinematic factor in the multiplet basis expansion in terms of
color-ordered amplitudes in the KK basis. To achieve this
in the recursion approach, we
recall that the $(n-1)$-gluon kinematic factor $A^{\alpha}(1,...,n)$ can
be written in terms of $(n-1)$-gluon color-ordered amplitudes $A(\sigma)$
(in the five-point case, the $(n-1)$-gluon result is given by
\eqref{4pt-kinematic-part}). Since the $(n-1)$-gluon
color-ordered amplitudes satisfy the KK relation \eqref{KK-relation},
we can apply KK relations to rewrite the $(n-1)$-gluon kinematic factor
in the multiplet basis in terms of $(n-1)$-gluon color-ordered amplitudes
of the form $A(1^-,\dots, \WH P^-_{i,n})$.
After substituting KK expressions for the $(n-1)$-gluon kinematic factors --
as well as the particular form of the MHV amplitudes \eqref{MHV-amplitude} --
into the $n$-gluon recursion expression (\eqref{5pt-kinematic-1} for five gluons),
we obtain the $n$-gluon kinematic factor expressed in terms of
combinations of $n$-gluon color-ordered MHV amplitudes of the
form $A(1^-,\dots,n^-)$.
Putting all this together, we arrive at the five-gluon result,
explicitly stated in \appref{sec:5g}, and electronically attached as an online resource.
Clearly, in the five-gluon case, the only non-vanishing helicity
configurations are the MHV and the $\overline{\mbox{MHV}}$ configuration,
making our result applicable to all relevant cases.

\subsection{The six-gluon amplitude}
\label{sec:six-point}

The six-gluon MHV amplitudes have been calculated analogously
using the electronically attached charge conjugation invariant
multiplet basis (see online resource). Concerning the basis, we remark that the
dimension of the all order vector space is reduced from 265
to 140 when keeping only conjugation
invariant linear combinations of vectors. Specializing to
$\Nc=3$ brings down the dimension further to 75.

We also note that although -- expressed in the KK basis --
there could naively be up to 4! spinor terms multiplying each basis vector,
on average only 8.5 contribute. By the scalar product approach in \secref{sec:comparison}, we find that the final expression for the MHV amplitude is also valid when we replace the KK basis for the MHV configuration by the basis for arbitrary helicity configurations. Thus we need not specify the helicity information in the final six gluon result.
The six-gluon basis, the radiation matrices and the resulting
amplitudes are electronically attached as online resources.

\section{Conclusion and outlook}
\label{sec:conclusion}

We have shown how BCFW recursion can be used for relating
higher point tree-level MHV gluon amplitudes to results for
fewer external legs.
To achieve this recursive decomposition we have utilized two
different strategies.

One option is to straightforwardly evaluate scalar products of
color factors in the multiplet bases with those in the DDM decomposition
(or in principle any other basis where recursion relations
are known). While this strategy benefits from being conceptually
simple, it will be competitive for multi-particle
processes only if a rapidly decreasing fraction of such
scalar products are non-zero, and if the contributing scalar
products can be identified and evaluated quickly.

Therefore we have shown how to derive $\Ng$-gluon
MHV amplitudes directly in the multiplet bases.
This requires the calculation
of ``radiation matrices'', describing the effect of radiating
one gluon from an $(\Ng-1)$-gluon basis vector, decomposed into the
$\Ng$-basis vectors. We have shown how to efficiently
calculate these matrices using birdtrack techniques, and
argued that the overall treatment of color structure can be
sped up significantly using multiplet bases.

While we do believe that the present paper is an important step
in the direction of achieving efficient multi-particle
amplitude calculations in multiplet bases, quite some
work remains before this becomes reality.
First of all, the results should be extended beyond MHV,
to processes with quarks, and preferably beyond leading
order. Secondly, it remains to 
efficiently implement multiplet bases and calculation of the
radiation matrices.

Finally, we remark that we have studied recursion using one
particular form of multiplet bases, corresponding to one
particular subgrouping of partons. It appears quite likely
that even  more efficient choices of multiplet bases can be made,
such that the radiation matrices are even more sparse and
the recursion can be achieved even quicker.

\subsection*{Acknowledgements}
We thank Johan Bijnens for useful comments on the manuscript.
This work was supported by the Swedish Research Council
(contract number 621-2012-27-44 and 621-2013-4287) and
in part by the MCnetITN FP7 Marie Curie Initial Training
Network, contract PITN-GA-2012-315877.
Y.D. would  like to acknowledge the supports from Erasmus Mundus Action 2, Project 9, the International Postdoctoral Exchange Fellowship Program of China (with Fudan University as the home university), the NSF of China Grant No. 11105118, China Postdoctoral Science Foundation No. 2013M530175 and the Fundamental Research Funds for the Central Universities of Fudan University No. 20520133169.

\appendix

\section{Vanishing of the (3,n-1) division }
\label{sec:3_n-1division}
Here we show the vanishing of the $(3,n-1)$ division in the BCFW recursion expression of the MHV amplitude ${\cal M}(g_1^-,g_2^+,\dots,g_n^-)$ for the $(1,n)$-shift, \eqref{eq:spinor_shift}. The proof here is similar to the proof used for color-ordered amplitudes which can be found, for example, in \cite{Elvang:2013cua}.
For the $(3,n-1)$ division, the left sub-amplitudes are three-gluon
$\overline{\text{MHV}}$ amplitudes, while the right sub-amplitudes are $(n-1)$-gluon MHV amplitudes.
Using the fact that the sub-amplitudes for fewer gluons always
can be expressed as linear combinations of color-ordered
amplitudes in the DDM basis, we find that one contribution to the kinematic
factor for this division has the form
\bea
i{\Spbb{i| -\WH P_{1,i}}^3\over\Spbb{\WH1\mid i}{\Spbb{\WH1\mid-\WH P_{1,i}}} }{i\over s_{1i}}i{\Spaa{\WH P_{1,i}\mid \WH n}^3\over\Spaa{\WH P_{1,i}\mid l}\dots \Spaa{k\mid \WH n}},~~\Label{(3,n-1)-shift}
\label{eq:3n-1}
\eea
where $l$ and $k$ are two arbitrary unshifted gluons in the right set,
and $\bket{-\WH P_{1,i}}=\pm i\bket{\WH P_{1,i}}$. (Since there are three $\bket{-\WH P_{1,i}}$'s in the numerator and one in the denominator, the sign $(\pm 1)$ does not appear in the final result.) In the above expression we have divided out a factor $\Spbb{i| -\WH P_{1i}}$ in the $\overline{\text{MHV}}$ amplitude and a factor $\Spaa{\WH P_{1,i}\mid \WH n} $ in the MHV amplitude.
The position of the pole for the above expression is
\bea
s_{1,i}(z)=\left(\Spbb{1|i}-z\Spbb{n|i}\right)\Spaa{i|1}=0 \Rightarrow z_{1,i}={\Spbb{1|i}\over \Spbb{n|i}}.
\eea
The numerator of  \eqref{eq:3n-1} then reads (up to a factor $i$)
\bea
\left(\Spbb{i\mid \WH P_{1,i}}\Spaa{\WH P_{1,i}\mid \WH n}\right)^3&=&\left(\Spbb{i\mid 1}\Spaa{1\mid \WH n}-z_{1,i}\Spbb{i\mid n}\Spaa{1\mid \WH n}\right)^3\nn
&=&\left(\Spbb{i\mid 1}-z_{1,i}\Spbb{i\mid n}\right)^3\Spaa{1\mid n}^3\nn
&=&0^3\Spaa{1\mid n}^3.
\eea
Also in the denominator, several factors vanish,
\bea
\Spbb{\WH1\mid i}=\Spbb{1\mid i}-z_{1,i}\Spbb{n\mid i}=0,
\eea
and
\bea
\Spbb{\WH1\mid\WH P_{1,i}}\Spaa{\WH P_{1,i}\mid l}
&=&\Spba{\WH1\mid \WH p_1+p_i\mid l}\nn
&=&-\Spaa{l\mid i}\left(\Spbb{1\mid i}-z_{1,i}\Spbb{n\mid i}\right)\nn
&=&0.
\eea
Thus, after dividing out a factor $\left(\Spbb{1\mid i}-z_{1,i}\Spbb{n\mid i}\right)^2$ in both numerator and denominator, the expression \eqref{(3,n-1)-shift} is proportional to
\bea
\Spbb{1\mid i}-z_{1,i}\Spbb{n\mid i}=0.
\eea
Therefore, the $(3,n-1)$ division always gives a vanishing contribution.

\section{Five-gluon multiplet basis, radiation matrices and MHV amplitudes}
\label{sec:5g}
We use the charge conjugation invariant orthonormal five-gluon multiplet basis
given below. As remarked in \secref{sec:color_recursion} we need, apart from
the representation labels $\alpha_i$, also a label
distinguishing the various vertices from each other.
For this reason the vectors carry labels
$\alpha_1,\alpha_2,\alpha_2'$ where, $\alpha_i$ contains
additional information about the vertex if needed.
Also, since our basis is charge conjugation invariant, the
representations $10$ and $\overline{10}$ only appear together,
referred to as $20$.
The five-gluon basis is

\begin{eqnarray}\label{eq:MultipletBasisVectors5g}
\Vec^{8a,1,1}_{g_1\, g_3\, g_5;\, g_2\, g_4}&=&
\frac{1}{ \left(\Nc^2-1\right) \sqrt{2 \Nc \TR}}
i f^{g_1 g_3 g_5} \delta^{g_2 g_4} \nonumber \\
\Vec^{1,8,8a}_{g_1\, g_3\, g_5;\, g_2\, g_4}&=&
\frac{1}{ \left(\Nc^2-1\right) \sqrt{2 \Nc \TR}}
\delta^{g_1 g_3} i  f^{g_2 g_5 g_4}  \nonumber \\
\Vec^{8s,8s,8a}_{g_1\, g_3\, g_5;\,g_2\, g_4}&=&
\frac{\sqrt{\Nc}}{2 \TR^{3/2} \left(\Nc^2-4\right)  \sqrt{2(\Nc^2-1)} }
 d^{g_1 g_3 i_1} d^{i_1 g_5 i_2 } if^{ i_2  g_4 g_2 } \nonumber \\
\Vec^{8s,8a,8s}_{g_1\, g_3\, g_5;\,g_2\, g_4}&=&
\frac{ \sqrt{\Nc} }{2 \TR^{3/2} \left(\Nc^2-4\right) \sqrt{2(\Nc^2-1)} }
d^{g_1 g_3 i_1} if^{i_1 g_5 i_2 } d^{i_2  g_4 g_2} \nonumber \\
\Vec^{8a,8s,8s}_{g_1\, g_3\, g_5;\,g_2\, g_4}&=&
\frac{\sqrt{\Nc}}{{ 2 \TR^{3/2}  \left(\Nc^2-4\right) \sqrt{2 (\Nc^2-1)} }}
if^{g_1 g_3 i_1} d^{i_1 g_5 i_2 } d^{i_2  g_4 g_2}   \nonumber \\
\Vec^{8a,8a,8a}_{g_1\, g_3\, g_5;\,g_2\, g_4}&=&
\frac{1}{2 (\Nc \TR)^{3/2} \sqrt{ 2 (\Nc^2-1)}  }
if^{g_1 g_3 i_1} if^{i_1 g_5 i_2 } if^{ i_2 g_4  g_2 }   \nonumber \\
\Vec^{27,8,8a}_{g_1\, g_3\, g_5;\,g_2\, g_4}&=&
\frac{\sqrt{2}}{\Nc^{3/2}   \sqrt{\TR \left(\Nc^2+2 \Nc-3\right) }}
\Proj^{27}{}_{ g_1\, g_3;\,i_1\, g_5 } if^{i_1 g_4 g_2 }   \nonumber \\
\Vec^{20,8,8s}_{g_1\, g_3\, g_5;\,g_2\, g_4}&=&
\frac{\sqrt{ \Nc }}{(\Nc^2-4) \sqrt{\TR \left(\Nc^2-1\right) }   }
\Proj^{10-\overline{10}} {}_{ g_1\, g_3; i_1\, g_5 } d^{i_1 g_4 g_2 }   \nonumber \\
\Vec^{20,8,8a}_{g_1\, g_3\, g_5;\,g_2\, g_4}&=&
\frac{1}{\sqrt{\Nc \TR \left(\Nc^4-5 \Nc^2+4\right) }}
\Proj^{10+\overline{10}}{}_{ g_1\, g_3; i_1\, g_5 } if^{i_1 g_4 g_2 }   \nonumber \\
\Vec^{0,8,8a}_{g_1\, g_3\, g_5;\,g_2\, g_4}&=&
\frac{\sqrt{2}}{\Nc^{3/2} \sqrt{\TR(\Nc^2-2 \Nc-3)} }
 \Proj^0 {}_{ g_1\, g_3;\, i_1\, g_5} i f^{ i_1 g_4 g_2}  \nonumber \\
\Vec^{8a,27,27}_{g_1\, g_3\, g_5;\,g_2\, g_4}&=&
\frac{\sqrt{2}}{\Nc^{3/2}  \sqrt{\TR \left(\Nc^2+2 \Nc-3\right) }}
i f^{g_1 g_3 i_1 }  \Proj^{27} {}_{ i_1\, g_5;\,g_2\, g_4 } \nonumber \\
\Vec^{27,27a,27}_{g_1\, g_3\, g_5;\,g_2\, g_4}&=&
\frac{2}{\Nc \sqrt{ \TR \left(\Nc^3+3 \Nc^2-\Nc-3\right)}}
\Proj^{27} {}_{ g_1\, g_3;\,i_1\,  i_3 } i f^{i_3 g_5 i_2}  \Proj^{27} {}_{ i_1\, i_2;\,g_2\, g_4 }   \nonumber \\
\Vec^{20,27,27}_{g_1\, g_3\, g_5;\,g_2\, g_4}&=&
\frac{2}{\sqrt{\Nc \TR \left(\Nc^4+\Nc^3-7 \Nc^2-\Nc+6\right) }}
 \Proj^{10+\overline{10}} {}_{ g_1\, g_3;\,i_1\, i_3 } i f^{ i_3 g_5 i_2 }  \Proj^{27} {}_{i_1\, i_2;\,g_2\, g_4  }  \nonumber \\
\Vec^{8s,20,20}_{g_1\, g_3\, g_5;\,g_2\, g_4}&=&
\frac{\sqrt{\Nc}} {(\Nc^2-4) \sqrt{\TR \left(\Nc^2-1\right)} }
d^{g_1 g_3 i_1}  \Proj^{10-\overline{10}} {}_{ i_1\, g_5;\,g_2\, g_4 } \nonumber \\
\Vec^{8a,20,20}_{g_1\, g_3\, g_5;\,g_2\, g_4}&=&
\frac{1}{\sqrt{\Nc \TR \left(\Nc^4-5 \Nc^2+4\right) }}
i f^{g_1 g_3 i_1}  \Proj^{10+\overline{10}} {}_{ i_1\, g_5;\,g_2\, g_4 } \nonumber \\
\Vec^{27,20,20}_{g_1\, g_3\, g_5;\,g_2\, g_4}&=&
\frac{2}{\sqrt{\Nc \TR \left(\Nc^4+\Nc^3-7 \Nc^2-\Nc+6\right) }}
\Proj^{27} {}_{ g_1\, g_3;\, i_1\, i_3 }  i f^{ i_3 g_5 i_2 }   \Proj^{10+\overline{10}} {}_{ i_1\, i_2;\,g_2\, g_4 } \nonumber \\
\Vec^{20,20f,20}_{g_1\, g_3\, g_5;\,g_2\, g_4}&=&
\frac{\sqrt{2}}{\sqrt{\Nc  \TR \left(\Nc^4-5 \Nc^2+4\right)}}
\Proj^{10+\overline{10}} {}_{ g_1\, g_3;\,i_1\, i_3 } if^{i_3 g_5 i_2}  \Proj^{10+\overline{10}} {}_{ i_1\, i_2;\,g_2\, g_4 }   \nonumber \\
\Vec^{20,20fd,20}_{g_1\, g_3\, g_5;\,g_2\, g_4}&=&
\frac{\sqrt{2 \Nc}}{\sqrt{\TR(\Nc^6-14 \Nc^4+49 \Nc^2-36)}}\left( \right.
\Proj^{10+\overline{10}}{}_{ g_1\, g_3;\,i_1\, i_3 }  d^{i_3 g_5 i_2}
\Proj^{10-\overline{10}} {}_{ i_1\, i_2;\,g_2\, g_4 }  \nonumber \\
&-& \frac{1}{\Nc} \Proj^{10+\overline{10}} {}_{ g_1\, g_3;\,i_1\, i_3 } i f^{i_3 g_5 i_2}  \Proj^{10+\overline{10}} {}_{ i_1\, i_2;\,g_2\, g_4} \left. \right) \nonumber \\
\Vec^{0,20,20}_{g_1\, g_3\, g_5;\,g_2\, g_4}&=&
\frac{2}{\sqrt{\Nc \TR  (\Nc^4-\Nc^3-7 \Nc^2+\Nc+6)} }
\Proj^{0} {}_{ g_1\, g_3;\, i_1\,  i_3 } if^{  i_3 g_5 i_2} \Proj^{10+\overline{10}} {}_{ i_1\, i_2;\, g_2\, g_4 } \nonumber \\
\Vec^{8a,0,0}_{g_1\, g_3\, g_5;\,g_2\, g_4}&=&
\frac{\sqrt{2}}{\Nc^{3/2} \sqrt{\TR(\Nc^2-2 \Nc-3)} }
i f^{g_1 g_3 i_1 }  \Proj^0 {}_{ i_1\, g_5;\, g_2\, g_4 }  \nonumber \\
\Vec^{20,0,0}_{g_1\, g_3\, g_5;\,g_2\, g_4}&=&
\frac{2}{\sqrt{ \Nc \TR (\Nc^4-\Nc^3-7 \Nc^2+\Nc+6)}}
\Proj^{10+\overline{10}} {}_{ g_1\, g_3;\, i_1\, i_3 } if^{i_3 g_5  i_2 }    \Proj^0 {}_{ i_1\, i_2;\,g_2\, g_4 } \nonumber \\
\Vec^{0,0a,0}_{g_1\, g_3\, g_5;\,g_2\, g_4}&=&
\frac{2}{\Nc \sqrt{\TR (\Nc^3-3 \Nc^2-\Nc+3)} }
\Proj^0 {}_{ g_1\, g_3;\, i_1\, i_3 } if^{i_3 g_5 i_2}  \Proj^0 {}_{  i_1\, i_2;\, g_2\, g_4 }.
\label{eq:5gBasis}
\end{eqnarray}

The radiation matrices for adding the gluon $g_5$ to the
four-gluon multiplet basis from \eqref{eq:4gNormalized}
expressed in the five-gluon basis
in \eqref{eq:5gBasis} are given by

\bea
\mathbf{T}_{2}=
\sqrt{\TR}\tiny{
\left[
\begin{array}{cccccc}
 0 & 0 & -\sqrt{\frac{2\Nc}{\Nc^2-1}} & 0 & 0 & 0 \\
 -\sqrt{2\Nc} & 0 & 0 & 0 & 0 & 0 \\
 0 & -\sqrt{\frac{\Nc}{2}} & 0 & 0 & 0 & 0 \\
 0 & \sqrt{\frac{\Nc}{2}} & 0 & 0 & 0 & 0 \\
 0 & 0 & -\sqrt{\frac{\Nc}{2}} & 0 & 0 & 0 \\
 0 & 0 & \sqrt{\frac{\Nc}{2}} & 0 & 0 & 0 \\
 0 & 0 & 0 & \sqrt{\frac{2}{\Nc}} & 0 & 0 \\
 0 & 0 & 0 & 0 & \sqrt{\frac{2 \Nc}{\Nc^2-4}} & 0 \\
 0 & 0 & 0 & 0 & 0 & 0 \\
 0 & 0 & 0 & 0 & 0 & -\sqrt{\frac{2}{\Nc}} \\
 0 & 0 & \sqrt{\frac{\Nc (\Nc+3)}{2(\Nc+1)}} & 0 & 0 & 0 \\
 0 & 0 & 0 & \sqrt{\Nc+1} & 0 & 0 \\
 0 & 0 & 0 & 0 & -\sqrt{\frac{\Nc (\Nc+3)}{2(\Nc+2)}} & 0 \\
 0 & \sqrt{\Nc} & 0 & 0 & 0 & 0 \\
 0 & 0 & 0 & 0 & 0 & 0 \\
 0 & 0 & 0 & -\sqrt{\frac{\Nc^2-\Nc-2}{\Nc}} & 0 & 0 \\
 0 & 0 & 0 & 0 & \sqrt{\Nc} & 0 \\
 0 & 0 & 0 & 0 & 0 & 0 \\
 0 & 0 & 0 & 0 & 0 & -\sqrt{\frac{\Nc^2+\Nc-2}{\Nc}} \\
 0 & 0 & -\sqrt{\frac{\Nc (\Nc-3)}{2(\Nc-1)}} & 0 & 0 & 0 \\
 0 & 0 & 0 & 0 & -\sqrt{\frac{\Nc (\Nc-3)}{2(\Nc-2)}} & 0 \\
 0 & 0 & 0 & 0 & 0 & \sqrt{\Nc-1} \\
\end{array}
\right]
}\;,
\eea

\bea\label{eq:RadiationMatrix_4g_to_5g_Gluon_3}
\mathbf{T}_{3}=
\sqrt{\TR} \tiny{
\left[
\begin{array}{cccccc}
 -\sqrt{2 \Nc}  & 0 & 0 & 0 & 0 & 0 \\
 0 & 0 & -\sqrt{\frac{2 \Nc}{\Nc^2-1}} & 0 & 0 & 0 \\
 0 & 0 & -\sqrt{\frac{\Nc}{2}} & 0 & 0 & 0 \\
 0 & -\sqrt{\frac{\Nc}{2}} & 0 & 0 & 0 & 0 \\
 0 & -\sqrt{\frac{\Nc}{2}} & 0 & 0 & 0 & 0 \\
 0 & 0 & -\sqrt{\frac{\Nc}{2}} & 0 & 0 & 0 \\
 0 & 0 & \sqrt{\frac{\Nc (\Nc+3)}{2(\Nc+1)}} & 0 & 0 & 0 \\
 0 & \sqrt{\Nc} & 0 & 0 & 0 & 0 \\
 0 & 0 & 0 & 0 & 0 & 0 \\
 0 & 0 & -\sqrt{\frac{\Nc (\Nc-3)}{2(\Nc-1)}} & 0 & 0 & 0 \\
 0 & 0 & 0 & \sqrt{\frac{2}{\Nc}} & 0 & 0 \\
 0 & 0 & 0 & -\sqrt{\Nc+1} & 0 & 0 \\
 0 & 0 & 0 & -\sqrt{\frac{\Nc^2-\Nc-2}{\Nc}} & 0 & 0 \\
 0 & 0 & 0 & 0 & \sqrt{\frac{2 \Nc}{\Nc^2-4}} & 0 \\
 0 & 0 & 0 & 0 & 0 & 0 \\
 0 & 0 & 0 & 0 & -\sqrt{\frac{\Nc (\Nc+3)}{2(\Nc+2)}} & 0 \\
 0 & 0 & 0 & 0 & -\sqrt{\Nc} & 0 \\
 0 & 0 & 0 & 0 & 0 & 0 \\
 0 & 0 & 0 & 0 & -\sqrt{\frac{\Nc (\Nc-3)}{2(\Nc-2)}} & 0 \\
 0 & 0 & 0 & 0 & 0 & -\sqrt{\frac{2}{\Nc}} \\
 0 & 0 & 0 & 0 & 0 & -\sqrt{\frac{\Nc^2+\Nc-2}{\Nc}} \\
 0 & 0 & 0 & 0 & 0 & -\sqrt{\Nc-1} \\
\end{array}
\right]
}
\eea
and
\bea
\mathbf{T}_{4}=
\sqrt{\TR}\tiny{\left[
\begin{array}{cccccc}
 0 & 0 & \sqrt{\frac{2 \Nc}{\Nc^2-1}} & 0 & 0 & 0 \\
 \sqrt{2 \Nc}  & 0 & 0 & 0 & 0 & 0 \\
 0 & \sqrt{\frac{\Nc}{2}} & 0 & 0 & 0 & 0 \\
 0 & \sqrt{\frac{\Nc}{2}} & 0 & 0 & 0 & 0 \\
 0 & 0 & \sqrt{\frac{\Nc}{2}} & 0 & 0 & 0 \\
 0 & 0 & \sqrt{\frac{\Nc}{2}} & 0 & 0 & 0 \\
 0 & 0 & 0 & -\sqrt{\frac{2}{\Nc}} & 0 & 0 \\
 0 & 0 & 0 & 0 & -\sqrt{\frac{2 \Nc}{\Nc^2-4}} & 0 \\
 0 & 0 & 0 & 0 & 0 & 0 \\
 0 & 0 & 0 & 0 & 0 & \sqrt{\frac{2}{\Nc}} \\
 0 & 0 & -\sqrt{\frac{\Nc (\Nc+3)}{2(\Nc+1)}} & 0 & 0 & 0 \\
 0 & 0 & 0 & \sqrt{\Nc+1} & 0 & 0 \\
 0 & 0 & 0 & 0 & \sqrt{\frac{\Nc (\Nc+3)}{2(\Nc+2)}} & 0 \\
 0 & -\sqrt{\Nc} & 0 & 0 & 0 & 0 \\
 0 & 0 & 0 & 0 & 0 & 0 \\
 0 & 0 & 0 & \sqrt{\frac{\Nc^2-\Nc-2}{\Nc}} & 0 & 0 \\
 0 & 0 & 0 & 0 & \sqrt{\Nc} & 0 \\
 0 & 0 & 0 & 0 & 0 & 0 \\
 0 & 0 & 0 & 0 & 0 & \sqrt{\frac{\Nc^2+\Nc-2}{\Nc}} \\
 0 & 0 & \sqrt{\frac{\Nc (\Nc-3)}{2(\Nc-1)}} & 0 & 0 & 0 \\
 0 & 0 & 0 & 0 & \sqrt{\frac{\Nc (\Nc-3)}{2(\Nc-2)}} & 0 \\
 0 & 0 & 0 & 0 & 0 & \sqrt{\Nc-1} \\
\end{array}
\right].
}
\eea

Using the above radiation matrices
(which can also be found among the electronic attachments available as online resource)
in \eqref{5pt-kinematic-1}
we arrive at the five-gluon multiplet basis result
\bea
 A^{\alpha}(1,2,3,4,5)=\Sl_{i=1}^6{\cal K}_i{ \cal C}_{i\alpha}{\cal N}_{\alpha},
\eea
where $\alpha=1,\dots, 22$, corresponding to the 22 vectors in
the five-gluon basis. Here  ${\cal K}$, ${\cal C}$ and ${\cal N}$ are defined by
\begin{equation}
{\cal K}=\left(
\begin{array}{cccccc}
 A(1,2,3,4,5),A(1,2,4,3,5),A(1,3,2,4,5),A(1,3,4,2,5), A(1,4,2,3,5),A(1,4,3,2,5)
\end{array}
\right),
\end{equation}
\begin{equation}
 {\cal C}=\left(
\begin{array}{cccccccccccccccccccccc}
 1 & 1 & 1 & 1 & 1 & 1 & 1 & 0 & 0 & 1 & 1 & 1 & 0 & 1 & 0 & 1 & 0 & 0 & 1 & 1 & 0 & 1 \\
 2 & \frac{1}{2} & 1 & 1 & 1 & 1 & -\frac{N_c}{2} & 1 & 0 & \frac{N_c}{2} & -\frac{2}{N_c} & 1 & 1 & 0 & 0 & 0 & 0 & 0 & 0 & \frac{2}{N_c}
& 1 & 1 \\
 2 & 0 & 0 & 0 & 2 & 2 & 0 & 0 & 0 & 0 & 2 & 0 & 0 & 0 & 0 & 0 & 0 & 0 & 0 & 2 & 0 & 0 \\
 2 & 0 & 0 & 0 & 2 & -2 & 0 & 0 & 0 & 0 & 2 & 0 & 0 & 0 & 0 & 0 & 0 & 0 & 0 & 2 & 0 & 0 \\
 2 & -\frac{1}{2} & -1 & 1 & 1 & -1 & \frac{N_c}{2} & 1 & 0 & -\frac{N_c}{2} & -\frac{2}{N_c} & 1 & 1 & 0 & 0 & 0 & 0 & 0 & 0 &
\frac{2}{N_c} & 1 & 1 \\
 1 & -1 & -1 & 1 & 1 & -1 & -1 & 0 & 0 & -1 & 1 & 1 & 0 & -1 & 0 & -1 & 0 & 0 & -1 & 1 & 0 & 1
\end{array}
\right)
\end{equation}
and\newpage
%
%
%
%
\bea
{\cal N}&=& ({\Nc} )^{3/2} \left(
\sqrt{2},
2 \sqrt{2},
\frac{\sqrt{\Nc^2-1}}{\sqrt{2}},
\frac{\sqrt{\Nc^2-1}}{\sqrt{2}},
\frac{\sqrt{\Nc^2-1}}{\sqrt{2}},
\frac{\sqrt{\Nc^2-1}}{\sqrt{2}},
\frac{\sqrt{2} \sqrt{\Nc^2+2 \Nc-3}}{\Nc},
-\sqrt{\Nc^2-1},
0,
\right.\nn
&&
\frac{\sqrt{2} \sqrt{\Nc^2-2 \Nc-3}}{\Nc},
-\frac{\sqrt{\Nc^2+2 \Nc-3}}{\sqrt{2}},
-\sqrt{\Nc^2+3 \Nc-\frac{3}{\Nc}-1},
-\frac{\sqrt{\Nc^4+\Nc^3-7 \Nc^2-\Nc+6}}{\Nc},
\nn
&&
-\sqrt{\Nc^2-1},
0,
-\frac{\sqrt{\Nc^4+\Nc^3-7 \Nc^2-\Nc+6}}{\Nc},
0,
0,
\frac{\sqrt{\Nc^4-\Nc^3-7 \Nc^2+\Nc+6}}{\Nc},
\frac{\sqrt{\Nc^2-2 \Nc-3}}{\sqrt{2}},\nn
&&\left.
\frac{\sqrt{\Nc^4-\Nc^3-7 \Nc^2+\Nc+6}}{\Nc},
\frac{\sqrt{\Nc^3-3 \Nc^2-\Nc +3}}{\sqrt{\Nc}}\right).
\eea

\bibliographystyle{JHEP}
\bibliography{Refs}

\private{

}

\end{document}